\documentclass[preprint]{elsarticle}
\usepackage[textwidth=19cm, textheight=24cm]{geometry}

\usepackage[utf8]{inputenc}
\usepackage[english]{babel}
\usepackage[T1]{fontenc}
\usepackage[hidelinks]{hyperref}
\usepackage{amsmath}
\usepackage{amssymb}
\usepackage{subcaption}
\usepackage{bm}
\usepackage{etoolbox}
\usepackage{siunitx}
\usepackage{algorithm}
\usepackage{algorithmicx}
\usepackage{algpseudocode}
\usepackage{setspace}
\usepackage{cancel}
\usepackage{nicefrac}
\usepackage{comment}
\usepackage{booktabs}

\hypersetup{pdfauthor=author}

\def\etal.{et\penalty50\ al.}

\makeatletter
\def\ps@pprintTitle{%
  \let\@oddhead\@empty
  \let\@evenhead\@empty
  \let\@oddfoot\@empty
  \let\@evenfoot\@oddfoot
}
\makeatother

\begin{document}

\begin{frontmatter}
\title{Sparse data assimilation for under-resolved large-eddy simulations}

\author[ifd]{Justin Plogmann}

\author[ifd]{Oliver Brenner}

\author[ifd]{Patrick Jenny\corref{corauthor}}
\cortext[corauthor]{Corresponding author: \url{jenny@ifd.mavt.ethz.ch}}

\affiliation[ifd]{%
    organization={Institute of Fluid Dynamics, ETH Zurich},%
    addressline={Sonneggstrasse~3},%
    city={Zürich},%
    postcode={CH-8092},%
    country={Switzerland}%
}

\begin{abstract}
The need for accurate and fast scale-resolving simulations of fluid flows, where turbulent dispersion is a crucial physical feature, is evident. Large-eddy simulations (LES) are computationally more affordable than direct numerical simulations, but their accuracy depends on sub-grid scale models and the quality of the computational mesh. In order to compensate related errors, a data assimilation approach for LES is devised in this work.

The presented method is based on variational assimilation of sparse time-averaged velocity reference data. Working with the time-averaged LES momentum equation allows to employ a stationary discrete adjoint method. Therefore, a stationary corrective force in the unsteady LES momentum equation is iteratively updated within the gradient-based optimization framework in conjunction with the adjoint gradient. After data assimilation, corrected anisotropic Reynolds stresses are inferred from the stationary corrective force. Ultimately, this corrective force that acts on the mean velocity is replaced by a term that scales the velocity fluctuations through nudging of the corrected anisotropic Reynolds stresses.

Efficacy of the proposed framework is demonstrated for turbulent flow over periodic hills and around a square cylinder. Coarse meshes are leveraged to further enhance the speed of the optimization procedure. Time- and spanwise-averaged velocity reference data from high-fidelity simulations is taken from the literature.

Our results demonstrate that adjoint-based assimilation of averaged velocity enables the optimization of the mean flow, vortex shedding frequency (i.\,e., Strouhal number), and anisotropic Reynolds stresses. This highlights the superiority of scale-resolving simulations such as LES over simulations based on the (unsteady) Reynolds-averaged equations. 

\end{abstract}


\end{frontmatter}


\section{Introduction}
\label{sec:intro}

The simulation of turbulent flows remains one of the major challenges in computational fluid dynamics (CFD). Direct numerical simulations (DNS) of complex flow problems at high Reynolds numbers are still unfeasible since the computational cost scales with $Re^3$. In turn, the accuracy of simulations based on the Reynolds-averaged Navier--Stokes (RANS) equations relies on model assumptions. Since all turbulent scales are modeled and none of the turbulence is resolved, the accuracy of RANS simulations highly depends on the choice of the turbulence model and the model parameters. For this reason, dissipation is often over-predicted in RANS simulations~\cite{wilcox_turbulence_2006}. The same applies to unsteady RANS (URANS) simulations~\cite{Durbin1995}. Nevertheless, RANS models are most commonly used for industrial flows, as their computational cost is low, but they are less suitable for flows where turbulent diffusion is important, e.\,g., in combustion, where species dispersion is crucial (e.\,g., in exhaust plumes in the wake of a vehicle~\cite{plogmann_urans_2023,Huang2020} or pollutant dispersion in urban flows~\cite{aristodemou_enhancing_2019}).

With increasing computational resources, large-eddy simulations (LES) are being used more and more in these areas. Large-eddy simulations involve solving the spatially filtered Navier--Stokes equations while the objective is to resolve the large turbulent structures and to model the effect of the small ones. Since only the smaller structures are directly subjected to modeling errors, LES provides a favorable compromise, particularly for free shear flows, as the computational expense is only moderately influenced by the Reynolds number (approximately proportional to $Re^{0.4}$) \cite{Chapman1979,Piomelli2002Wall-LayerSimulations}. When it comes to wall-resolved LES, however, the computational cost scales approximately with $Re^{1.8}$ \cite{ferziger_computational_2020,Pope2000}. Due to the high cost of LES for applications with wall turbulence, hybrid LES/RANS methods have received considerable attention \cite{Frohlich2008}. There, LES is typically supported by RANS simulations in regions where the LES is under-resolved (e.\,g.,~\cite{xiao12,spalart1997comments}). More recently, the Pseudo-DNS method, a multiscale framework that combines offline DNS-based databases of representative volume elements with coarse-scale flow simulations to efficiently model turbulent incompressible flows was introduced~\cite{idelsohn_p-dns_2024,idelsohn_pseudo-dns_2020,gimenez_multiscale_2025,larreteguy_datadriven_2023}.

In this work, however, we aim to enhance LES by data assimilation (DA) in order to reduce errors that are introduced by, e.\,g., the turbulence model or a coarse computational mesh~\cite{duraisamy19,singh16}. In general, there exist two main approaches in DA. In statistical DA, e.\,g., based on the ensemble Kalman filter (EnKF)~\cite{evensen_ensemble_2003}, observations are incorporated into the model one at a time as they become available. This approach sequentially updates the model's state estimate based on the current observations. For instance, in weather forecasting, sequential DA allows for continuous integration of observations from satellites, weather stations, and other sources to improve short-term predictions. Variational DA, on the other hand, formulates the assimilation problem as an optimization task, seeking the best model state that fits both the observations and the model dynamics. It involves minimizing a cost function that measures the misfit between model predictions and observations subject to constraints and regularization terms~\cite{asch16,evensen_data_2022}.

As an alternative approach, physics-informed neural networks (PINN) have been introduced by Raissi~\etal.~\cite{raissi18} to solve physical problems with neural networks instead of using conventional numerical partial differential equation (PDE) solvers. Patel~\etal.~\cite{patel_turbulence_2024} compared a variational DA framework with a PINN-based approach for RANS simulations over periodic hills and concluded that mean flow reconstruction could be achieved using either approach with a similar degree of accuracy. However, the overall computational cost of the PINN-based framework was higher. For more complex fluid flows, e.\,g. in industrial applications, Zhao~\etal.~\cite{10.1063/5.0226562} pointed out that PINN often suffer from slow and unstable convergence when solving high-dimensional PDEs and struggle with high-Reynolds-number turbulence.

DA can also be categorized into stationary and dynamic~\cite{asch16}. Dynamic DA approaches like the EnKF or four-dimensional variational (4DVar) DA incur huge computational costs since the flow dynamics cannot be neglected over a time interval during which observations are assimilated. EnKF have been extensively applied in the context of meteorology, and more recently also in the field of fluid mechanics~\cite{MELDI2017207,zhang_ensemble_2022}. Different attempts to reduce the computational cost of 4DVar DA for turbulent flows were also made (cf.~\cite{chandramouli_4d_2020,he_four-dimensional_2024,li_unsteady_2023}). This allowed to assimilate time-resolved data in LES, but the computational cost associated with these dynamic DA approaches is still very high, and time-resolved data is required.

Recent works have applied stationary DA to optimize simulations of steady-state flows using sparse data by performing three-dimensional variational (3DVar) DA, enabling the reconstruction of mean flow at a low computational cost (e.\,g.,~\cite{foures14,brenner22,brenner_variational_2024,li22,patel_turbulence_2024,symon_data_2017,franceschini_mean-flow_2020,symon_mean_2020,ghosh_robust_2024}. In addition, Brenner \etal.~\cite{brenner_variational_2024} compared different DA approaches for canonical stationary two-dimensional turbulent flow problems. Their results suggest that a divergence-free force as DA parameter achieves the optimization goal more efficiently compared to applying DA for obtaining the eddy viscosity, or a field modifying the eddy viscosity, directly.

An extension of 3DVar DA for unsteady flows was presented in Plogmann~\etal.~\cite{plogmann23}. There, time-averaging the URANS momentum equation was introduced to construct a stationary adjoint equation so that sparse time-averaged velocity reference data could ultimately be assimilated. This allowed to optimize a stationary (and divergence-free) corrective forcing term in the (unsteady) URANS momentum equation, which yielded mean flow reconstruction and an improved vortex shedding frequency of different turbulent wake flows at a low computational cost. An extension of their method was presented in \cite{plogmann24} for flows with time-periodic statistics. Central to their proposed methodology is the introduction of a corrective, divergence-free, and unsteady forcing term derived from a Fourier series expansion into the unsteady momentum equation. This term allows the tuning of stationary parameters across different Fourier modes, whereby the flow dynamics could be further improved.

\subsection{Objective and novelty of the present work}

In the present study, we aim to assimilate sparse time-averaged velocity reference data (obtained from high-fidelity LES and DNS) into coarse LES. Therefore, the framework from~\cite{plogmann23} is adapted, but LES serve as the forward problem instead of URANS simulations. We investigate turbulent flows over a periodic hill at a Reynolds number of 10595 and around a square cylinder at a Reynolds number of 22000. Notably, after optimization, the mean flow is in better agreement with the reference. The corrective, stationary, and divergence-free force, obtained from DA, is used to infer the corrected anisotropic part of the Reynolds stresses and marks a major extension to the DA framework developed in~\cite{plogmann23}. Ultimately, the stationary force for mean flow reconstruction is substituted by a velocity fluctuation force term, which additionally accounts for the corrected anisotropic part of the Reynolds stress tensor.

Furthermore, our DA framework comes with multiple advantages. Only time-averaged data is required, even though the simulations are unsteady. Additionally, the cost of the employed 3DVar DA scheme is relatively low due to our efficient semi-analytical approach for the computation of the cost function gradient within the discrete adjoint method (cf.~\ref{app:Computational cost discrete adjoint method}). The presented DA approach is based on the discrete adjoint method that itself relies on the explicit availability of a system matrix of the linearized and discretized forward problem equations. For meshfree methods that do provide such a matrix, the authors expect the presented DA framework to be applicable. However, some meshfree approaches, e.g. smoothed particle hydrodynamics, feature dense system matrices, which would lead to a dense transpose matrix in the adjoint system. This is expected to incur greater computational cost and maybe even convergence problems for the numerical solution of the adjoint system of equations. 

The remainder of the paper is organized as follows. The data assimilation framework is introduced in Sec.~\ref{sec:methods}, the anisotropic Reynolds stress reconstruction is presented in Sec.~\ref{sec:Anisotropic Reynolds stress reconstruction from corrective force}, and the velocity fluctuation scaling approach is discussed in Sec.~\ref{sec:Replacing corrective forcing with velocity fluctuation nudging}. Finally, in Sec.~\ref{sec:conclusion}, the work is summarized, and future developments are suggested.


\section{Time-averaged velocity data assimilation}
\label{sec:methods}

In this section, the data assimilation approach and results are presented for flows over periodic hills and around a square cylinder.

\subsection{Problem statement}
\label{sec: Problem statement}

\subsubsection{Filtered Navier--Stokes equations}
\label{sec:Unsteady Reynolds-averaged Navier--Stokes equations}

In the following, the turbulent flow of an incompressible Newtonian fluid is considered. Applying a low-pass filter to the Navier--Stokes equations yields the governing equations for LES. Thus, any instantaneous field $\xi$ is split into a filtered part $\bar{\xi}$ and a residual part $\xi'$. The LES governing equations are written as
\begin{equation}
\label{eq:continuity}
\frac{\partial \bar{u}_j}{\partial x_j} = 0
\end{equation}
and
\begin{equation}
    \label{eq:urans_momentum_no_assumption}
    \frac{\partial \bar{u}_{i}}{\partial t}
    +
    \frac{\partial \bar{u}_{i} \bar{u}_{j}}{\partial x_{j}}
    +
    \frac{\partial}{\partial x_{i}}
    \left[
        \frac{\bar{p}}{\rho}
    \right]
    -
    \frac{\partial}{\partial x_{j}}
    \left[
        2 \nu \bar{S}_{ij}
    \right]
    +
    \frac{\partial \tau_{ij}^{r}}{\partial x_{j}}
    = 0
\end{equation}
with constant density $\rho$ and the filtered rate-of-strain tensor
\begin{equation}
    \bar{S}_{ij}
    =
    \frac{1}{2}
    \left(
        \frac{\partial \bar{u}_{i}}{\partial x_{j}}
        +
        \frac{\partial \bar{u}_{j}}{\partial x_{i}}
    \right)
    \, .
\end{equation}

The residual stresses
\begin{equation}
\label{eq:sgs stresses}
    \tau_{ij}^{r} = \overline{u_i u_j} - \bar{u}_i\bar{u}_j
    \, 
\end{equation}
are modeled using the Boussinesq hypothesis, i.\,e.,
\begin{equation}
    \tau_{ij}^{r} \approx \frac{2}{3}k_\mathrm{sgs} \delta_{ij} - 2 \nu_\mathrm{sgs} \bar{S}_{ij}
\end{equation}
with the sub-grid scale (SGS) viscosity $\nu_\mathrm{sgs}$ and the isotropic part of the residual stress tensor. The SGS turbulent kinetic energy (TKE) is defined as 
\begin{equation}
    k_\mathrm{sgs} = \frac{1}{2} \tau^r_{ii}
    \, .
\end{equation}

\subsubsection{Data assimilation parameter}
\label{sec:Data assimilation parameter}

To account for discrepancies in the divergence of the residual stresses, we introduce a stationary corrective force $\bm F$ such that
\begin{equation}
    \frac{\partial \tau_{ij}^{r}}{\partial x_{j}}
    \approx
    \frac{\partial}{\partial x_{j}}
    \left(
        \frac{2}{3}k_\mathrm{sgs}\delta_{ij}
        -
        2\nu_\mathrm{sgs}\bar{S}_{ij}
    \right)
    -
    F_{i}
    \, ,
\end{equation}
which is then subjected to a Stokes--Helmholtz decomposition, similarly done in~\cite{foures14,perot_turbulence_1999,li22,patel_turbulence_2024}, i.\,e.,
\begin{equation}
    F_{i}
    =
    -\frac{\partial \phi}{\partial x_{i}}
    +
    \epsilon_{ijk} \frac{\partial \psi_{k}}{\partial x_{j}}
\end{equation}
with the scalar potential $\phi$, the vector potential $\bm \psi$, and the Levi-Civita symbol $\epsilon_{ijk}$.

For the modeled LES momentum equation this yields
\begin{equation}
    \label{eq:rans_momentum}
    \frac{\partial \bar{u}_{i}}{\partial t}
    +
    \frac{\partial \bar{u}_{i} \bar{u}_{j}}{\partial x_{j}}
    +
    \frac{\partial p^{*}}{\partial x_{i}}
    -
    \frac{\partial}{\partial x_{j}}
    \left[
        2\nu_{\mathrm{eff}}\bar{S}_{ij}
    \right]
    -
    \epsilon_{ijk} \frac{\partial \psi_{k}}{\partial x_{j}}
    =
    0
\end{equation}
with the effective viscosity
\begin{equation}
    \nu_{\mathrm{eff}}
    =
    \nu
    +
    \nu_\mathrm{sgs}
    \, ,
\end{equation}
and where the filtered pressure $\bar{p}$, the residual TKE $k_\mathrm{sgs}$, and the scalar potential $\phi$ are absorbed into the modified pressure as
\begin{equation}
    \label{eq:rans_pressure_mod}
    p^{*}
    =
    \frac{\bar{p}}{\rho}
    +
    \frac{2}{3} k_\mathrm{sgs}
    +
    \phi
    \ .
\end{equation}

Out of convenience, the modified pressure $p^{*}$ is denoted as $\bar{p}$ from now on. The data assimilation acts directly on $\psi_{k}$, that is, no additional equations need to be solved for $\phi$ and $\psi_{k}$. 

\subsubsection{Temporal averaging of the LES equations}
\label{sec:Temporal averaging of URANS equations}

To leverage the discrete adjoint method for stationary flows from \cite{brenner_variational_2024} and apply it to unsteady flow problems, temporal averaging is introduced. Therefore, a filtered quantity $\bar{\xi}$ is split into a temporal average $\left<\cdot\right>$ and corresponding fluctuation $\left(\cdot\right)''$ as
\begin{equation}
    \bar{\xi}
    = \left<\bar{\xi}\right>
    + \bar{\xi}''
    \, .
\end{equation}

A detailed derivation of time-averaging the momentum equation is given in~\cite{plogmann23}. With all terms expanded and rearranged, we obtain
\begin{equation}
\label{eq:urans_time_average}
    R_{\langle \bar{u}_i \rangle}
    =
    \frac{\partial \left<\bar{u}_{i}\right> \left<\bar{u}_{j}\right>}{\partial x_{j}}
     + \frac{\partial \left<\bar{p}\right>}{\partial x_{i}}
    - \frac{\partial}{\partial x_{j}}
        \left(
            2 \left(\nu + \left<\nu_\mathrm{sgs}\right>\right) \left<\bar{S}_{ij} \right>
        \right)
        - \underbrace{\frac{\partial}{\partial x_{j}}
        \left(
            2 \left<\nu''_\mathrm{sgs} \bar{S}_{ij}'' \right>
        \right)
     + \frac{\partial \left<\bar{u}''_{i} \bar{u}''_{j}\right>}{\partial x_{j}}}_{\mathrm{additional~stress~terms}}
     - \epsilon_{ijk} \frac{\partial \psi_{k}}{\partial x_{j}}
     = 0
\end{equation}
for the LES momentum equation residual.

The structure of Eq.~\eqref{eq:urans_time_average} is very similar to the modeled LES momentum equation~\eqref{eq:rans_momentum}, except that it describes a stationary state and that it features two additional terms due to the time-averaging process, which are treated explicitly in the adjoint problem discretization. The discrete adjoint method is now applied to the time-averaged LES equation \eqref{eq:urans_time_average}. 

To ensure converged averaged properties of the flow fields, which are needed in conjunction with their corresponding fluctuations to construct Eq.~\eqref{eq:urans_time_average}, we introduce a global measure $C_U$ for the change in an averaged field (cf. Eq.~\eqref{eq:averaging convergence criterion}). The criterion is set such that the averaged quantities converge well, but the number of time steps needed in the forward solution is kept as small as possible to minimize the required computational cost. Therefore, the criterion is case-dependent and is set individually for each case as discussed in sections \ref{sec:flow over periodic hills} and \ref{sec:flow around square cylinder}.

\subsection{Forward problem formulation}
\label{sec:forward problem formulation}

The forward problem solution consists of solving the momentum equation for $\bar{\bm u}$ and the pressure equation for $\bar{p}$. Hence, the time-dependent solution vector of the forward problem is defined as
\begin{equation}
    \begin{bmatrix}
        \bar{\bm u} \\
        \bar{\bm p}
    \end{bmatrix}
    =
    \begin{bmatrix}
         \bar{\bm u}_{x} \\
         \bar{\bm u}_{y} \\
         \bar{\bm u}_{z} \\
         \bar{\bm p}
    \end{bmatrix}
    \, ,
\end{equation}
and is obtained via the PISO algorithm discussed in section~\ref{sec:Implementation}. Note that $\bar{\bm u}_{x}$, $\bar{\bm u}_{y}$, $\bar{\bm u}_{z}$, and $\bar{\bm p}$ are vectors of length $n$, where $n$ is the number of grid cells.

Throughout this paper, indices 1, 2, and 3 correspond to spatial coordinates $x$, $y$, and $z$, respectively, such that $x_1$, $x_2$, and $x_3$ are equivalent to $x$, $y$, and $z$, and tensor components with repeated indices (e.\,g., $a_{11}$) represent the corresponding coordinate-specific elements (e.\,g., $a_{xx}$).

\subsection{Time-averaged forward problem formulation}
\label{sec:Time-averaged forward problem formulation}

While solving the forward problem, time-averaging is performed to obtain $\langle\bar{\bm u}\rangle$, $\langle\bar{\bm p}\rangle$, $\langle\bm \nu_\mathrm{sgs}\rangle$, and the additional terms in Eq.~\eqref{eq:urans_time_average}. The time-averaged forward system of equations is linearized in a \emph{coupled} manner as
\begin{equation}
    \label{eq:coupled_residual}
    R\left(\bm \psi, \bm U\right)
    =
    \begin{bmatrix}
        R_{\langle\bar{u}\rangle}\left(\bm\psi, \bm U\right) \\
        R_{\langle\bar{p}\rangle}\left(\bm\psi, \bm U\right)
    \end{bmatrix}
    =
    \begin{bmatrix}
        \mathbf{A}_{\langle\bar{u}\rangle\langle\bar{u}\rangle} & \mathbf{A}_{\langle\bar{u}\rangle\langle\bar{p}\rangle} \\
        \mathbf{A}_{\langle\bar{p}\rangle\langle\bar{u}\rangle} & \mathbf{A}_{\langle\bar{p}\rangle\langle\bar{p}\rangle}
    \end{bmatrix}
    \begin{bmatrix}
        \langle\bar{\bm u}\rangle \\
        \langle\bar{\bm p}\rangle
    \end{bmatrix}
    -
    \begin{bmatrix}
        \bm b_{\langle\bar{u}\rangle} \\
        \bm b_{\langle\bar{p}\rangle}
    \end{bmatrix}
    =
    \mathbf{A}_{U}
    \bm U
    -
    \bm b_{U}
    =
    0
    \, ,
\end{equation}
where $\bm \psi$ represents the parameters used for data assimilation and 
\begin{equation}
    \bm U
    =
    \begin{bmatrix}
        \langle\bar{\bm u}\rangle \\
        \langle\bar{\bm p}\rangle
    \end{bmatrix}
    =
    \begin{bmatrix}
         \langle\bar{\bm u}_{x}\rangle \\
         \langle\bar{\bm u}_{y}\rangle \\
         \langle\bar{\bm u}_{z}\rangle \\
         \langle\bar{\bm p}\rangle
    \end{bmatrix}
    \, ;
\end{equation}
again, $\langle\bar{\bm u}_{x}\rangle$, $\langle\bar{\bm u}_{y}\rangle$, $\langle\bar{\bm u}_{z}\rangle$, and $\langle\bar{\bm p}\rangle$ are vectors of length $n$. The system matrix $\mathbf{A}_{U}$ of the \emph{coupled} linear system of time-averaged equations is composed of sub-matrices that describe the implicit contributions.

\subsection{Inverse problem formulation}
\label{sec:Data assimilation problem}

To measure the misfit between the LES model output and the existing reference data, the scalar cost function $f$ is introduced. It consists of a regularization function $f_{\psi}$ and a discrepancy contribution $f_{U}$, i.\,e.,
\begin{equation}
    \label{eq:cost_function}
    f\left(\bm\psi, \bm U\right)
    =
    f_{\psi}\left(\bm \psi\right)
    +
    f_{U}\left(\bm U\right)
    \, ,
\end{equation}
and the optimization problem
\begin{subequations}
    \begin{alignat}{2}
        \label{eq:minimization_problem}
        & \!\min_{\psi}     & \quad & f\left(\bm\psi, \bm U\right) \\
        \label{eq:minimization_problem_constr_1}
        & \text{subject to} &       & \bm R\left(\bm\psi, \bm U\right) = 0\ 
    \end{alignat}
\end{subequations}
describes the data assimilation procedure, where $\bm \psi$ is the parameter vector to be optimized and $\bm U$ is the time-averaged forward problem solution. Due to the non-linearity, a non-linear optimization solver is used, but without regularization there is no assurance that the solution is unique~\cite{foures14}. Therefore, regularization is introduced to reduce the ambiguity (see Sec.~\ref{sec:Cost function and regularization}).

\subsubsection{Discrete adjoint method}
\label{sec:Discrete adjoint method}

The cost function gradient is derived by defining a Lagrangian
\begin{equation}
    \label{eq:lagrangian}
    \mathcal{L}\left(\bm \psi,\bm U\right)
    =
    f\left(\bm\psi,\bm U\right)
    -
    \bm\lambda^{T} \bm R\left(\bm\psi,\bm U\right)
\end{equation}
with the spatially varying Lagrange multipliers
\begin{equation}
    \label{eq:lambda}
    \bm\lambda
    =
    \begin{bmatrix}
        \bm \lambda_{\langle\bar{u}\rangle} \\
        \bm \lambda_{\langle\bar{p}\rangle}
    \end{bmatrix}
    =
    \begin{bmatrix}
        \bm \lambda_{\langle\bar{u}_{x}\rangle} \\
        \bm \lambda_{\langle\bar{u}_{y}\rangle} \\
        \bm \lambda_{\langle\bar{u}_{z}\rangle} \\
        \bm \lambda_{\langle\bar{p}\rangle}
    \end{bmatrix}
    \, ,
\end{equation}
where $\bm \lambda_{\langle\bar{u}_{x}\rangle}$, $\bm\lambda_{\langle\bar{u}_{y}\rangle}$, $\bm\lambda_{\langle\bar{u}_{z}\rangle}$, and $\bm\lambda_{\langle\bar{p}\rangle}$ are vectors of length $n$. The cost function gradient with respect to the parameters $\bm\psi$ is derived as
\begin{equation}
    \label{eq:adjoint_gradient}
    \frac{\mathrm{d} f}{\mathrm{d} \bm\psi}
    =
    \frac{\partial f_{\psi}}{\partial \bm\psi}
    -
    \bm\lambda^{T} \frac{\partial \bm R}{\partial \bm\psi}
    \, ,
\end{equation}
and following Brenner~\etal.~\cite{brenner22}, $\bm\lambda$ is obtained by solving the adjoint equation
\begin{equation}
    \label{eq:coupled_adjoint}
    \left( \frac{\partial \bm R}{\partial \bm U} \right)^{T} \bm\lambda
    \approx
    \mathbf{A}_{U}^{T} \, \bm\lambda
    =
    \frac{\partial f_{U}}{\partial \bm U}^{T}
    \, ,
\end{equation}
where the right-hand side is analytically derived from the cost function and thus comes at low computational cost.
The cost of solving this system for $\bm\lambda$ is negligible compared to that for solving the forward problem, and it is independent of the number of parameters $\bm \psi$ ($3n$).

Applied to the time-averaged LES forward problem with residual Eq.~\eqref{eq:urans_time_average} and cost function $f_U$, the \emph{coupled} adjoint system of equations~\eqref{eq:coupled_adjoint}, reads
\begin{equation}
    \renewcommand\arraystretch{1.2}
    \begin{bmatrix}
        \mathbf{A}_{\langle\bar{u}\rangle\langle\bar{u}\rangle}^{T} & \mathbf{A}_{\langle\bar{p}\rangle\langle\bar{u}\rangle}^{T} \\
        \mathbf{A}_{\langle\bar{u}\rangle\langle\bar{p}\rangle}^{T} & \mathbf{A}_{\langle\bar{p}\rangle\langle\bar{p}\rangle}^{T}
    \end{bmatrix}
    \begin{bmatrix}
        \bm\lambda_{\langle\bar{u}\rangle} \\
        \bm\lambda_{\langle\bar{p}\rangle}
    \end{bmatrix}
    =
    \begin{bmatrix}
        \frac{\partial f_{\langle\bar{u}\rangle}}{\partial \langle\bar{\bm u}\rangle}^{T} \\
        \frac{\partial f_{\langle\bar{p}\rangle}}{\partial \langle\bar{\bm p}\rangle}^{T}
    \end{bmatrix}
    \, .
\end{equation}

For the evaluation of the adjoint gradient in Eq.~\eqref{eq:adjoint_gradient}, the derivative of $\bm R$ with respect to parameter $\bm \psi$ is needed. Thus, the forward problem ($\bm R$) is numerically linearized with respect to parameter $\bm \psi$ in \emph{OpenFOAM} as
\begin{equation}
    \bm R
    =
    \mathbf{A}_{\psi} \, \bm\psi
    -
     \bm b_{\psi}
    =
    0
    \, ,
\end{equation}
such that the derivative of $\bm R$ with respect to parameter $\bm \psi$ reads
\begin{equation}
\label{eq:dRdPsi}
    \frac{\partial \bm R}{\partial \bm \psi}
    =
    \frac{\partial}{\partial \bm \psi}
    \left[
        \mathbf{A}_{\psi} \, \bm\psi
        -
        \bm b_{\psi}
    \right]
    =
    \mathbf{A}_{\psi}
    \, .
\end{equation}

Solving Eq.~\eqref{eq:coupled_adjoint} for $\bm\lambda$ using Eq.~\eqref{eq:dRdPsi} results in
\begin{equation}
    \label{eq:rans_adjoint_gradient}
    \frac{\mathrm{d} f}{\mathrm{d} \bm\psi}
    =
    \frac{\partial f_{\psi}}{\partial \bm\psi}
    -
    \bm\lambda_{\langle\bar{u}_{x}\rangle}^{T} \mathbf{A}_{\langle\bar{u}_{x}\rangle,\psi}
    -
    \bm\lambda_{\langle\bar{u}_{y}\rangle}^{T} \mathbf{A}_{\langle\bar{u}_{y}\rangle,\psi}
    -
    \underbrace{\bm\lambda_{\langle\bar{u}_{z}\rangle}^{T} \mathbf{A}_{\langle\bar{u}_{z}\rangle,\psi}}_{= \ 0 \ \mathrm{if} \ 2\mathrm{D}}
    -
    \underbrace{\bm\lambda_{\langle\bar{p}\rangle}^{T} \mathbf{A}_{\langle\bar{p}\rangle,\psi}}_{= \ 0}
\end{equation}
for the discrete adjoint gradient. Due to the divergence-free property of the forcing term, there is no contribution of parameter $\psi_{k}$ to the pressure equation. The derivative of the pressure residual with respect to the parameter is thus zero, and there is no contribution of the adjoint pressure $\bm\lambda_{\langle\bar{p}\rangle}$ to the adjoint gradient.

\subsubsection{Cost function and regularization}
\label{sec:Cost function and regularization}

The discrepancy part of the cost function measures agreement of the time-averaged forward problem solution $\bm U$ with the reference data $\bm U^{\mathrm{ref}}$.
In the presented application, only time- and spanwise-averaged velocity data is assimilated, i.\,e.,
\begin{equation}
    \label{eq:discrepancy}
    f_{U}\left(\bm U\right)
    =
    \sum\limits_{j\in\mathcal{R}} \left[
        \sum\limits_{k\in\left\{x,y,z\right\}}
        \left(
            \langle\bar{u}_{k,j}\rangle
            -
             \langle\bar{u}_{k,j}^{\mathrm{ref}}\rangle
        \right)^{2}
    \right]
    \ ,
\end{equation}
where $\mathcal{R}$ is the list of reference cell indices $j$. To reduce the ambiguity of the inverse problem, $L_2$ regularization is chosen. Therefore, regularization is applied to the (divergence-free) corrective forcing term. The reason for choosing this approach is that the forcing term is introduced to correct the mean flow but not the fluctuations of the velocity. In terms of time-averaged velocity, the initial LES is supposed to give adequate results relatively close to the time-averaged velocity reference data. It is, therefore, in order to avoid the suppression or damping of velocity fluctuations, that only a minimal correction is sought.

Hence, a function of the form
\begin{equation}
    \label{eq:regularization}
    f_{\bm\psi}\left(\psi\right)
    =
    C^{\mathrm{reg}}
    \lVert
        \nabla \times \bm\psi
    \rVert^{2}_{2}
\end{equation}
with hyperparameter $C^{\mathrm{reg}}$ is used to punish strong peaks in the corrective forcing term field. An appropriate hyperparameter is found when cost and test functions (see Sec.~\ref{sec:test function}) decrease during optimization.

\subsubsection{Test function}
\label{sec:test function}

In addition to the cost function we introduce the test function
\begin{equation}
    \label{eq:test function}
    f^\mathrm{test}
    =
    \frac{
        1
    }{
        V^{\mathrm{test}}
    }
    \sum\limits_{j\in\mathcal{T}} \left[
        \sum\limits_{k\in\left\{x,y,z\right\}}
        \left(
            \langle\bar{u}_{k,j}\rangle
            -
             \langle\bar{u}_{k,j}^{\mathrm{ref}}\rangle
        \right)^{2}
    V_j
    \right]
    \ ,
\end{equation}
where $\mathcal{T}$ is the list of test cell indices $j$, $V_{j}$ the volume of cell $j$, and $V^{\mathrm{test}}$ the total volume of all test cells. The test data points are all remaining points in the domain that are not defined as reference points. This concept is often dealt with in machine learning, where training, validation, and test data are considered. 

\subsection{Optimization}
\label{sec:Optimization}

The gradient-based fixed step size optimization algorithm is employed to perform updates of optimization parameter $\psi$ from iteration step $\left(n\right)$ to $\left(n+1\right)$ as
\begin{equation}
    \psi_{i}^{\left(n+1\right)}
    =
    \psi_{i}^{\left(n\right)}
    -
    \eta
    \frac{\mathrm{d} f}{\mathrm{d}\psi_{i}}^{\left(n\right)}
\end{equation}
with the fixed step size $\eta$, which is case dependent.

\subsection{Reference data}
\label{sec:Reference data}

Data assimilation was performed using public online reference data from the literature~\cite{gloerfelt19,trias_turbulent_2015}, in particular, spanwise and temporally averaged LES or DNS velocity data.

The \emph{SciPy} library in \emph{Python} offers the \emph{griddata} interpolation function, which is utilized for unstructured data interpolation on \emph{OpenFOAM} meshes to map reference data onto cell center locations. Specifically, the nearest neighbor method is utilized for this interpolation. To differentiate between reference and test data sets within these fields, two indicator fields are introduced, marking cells designated for testing and reference purposes, respectively.

We would like to emphasize at this point that the method is not limited to the use of DNS or LES data. Experimental data that has been sparsely measured (e.\,g., through point measurements) can also be used. However, to evaluate the success of data assimilation, the measurement points should also be divided into reference data points and test data points. This allows for appropriate tuning of the hyperparameters (i.\,e. optimization step size $\eta$ and regularization parameter $C^{\mathrm{reg}}$). The corresponding DA results then also show that the time-averaged velocity not only matches the experimental values in the assimilated reference data points, but also fits better in the test data points.

\subsection{Implementation}
\label{sec:Implementation}

The versions \textit{OpenFOAM-v1912}~\cite{of1912} and \textit{foam-extend-5.0}~\cite{fe50} of the open-source field operation and manipulation platform, \textit{OpenFOAM}, known for its computational fluid dynamics (CFD) solvers, are utilized for the forward and adjoint problems, respectively, leveraging the platform's pre-existing solvers and diverse capabilities. The approach adopted for solving the forward problem involves the PISO algorithm, and a fully \emph{coupled} solution process is chosen for the adjoint problem. Notably, the \textit{pisoFoam} and \textit{transientFoam} solvers were extended, respectively.

The computational meshes were created using \textit{blockMesh}, and the cell size is decreasing toward the solid walls to capture certain flow features well enough. However, the resolution in the wall-normal direction was still chosen quite coarse in order to allow for larger time steps, which in turn drastically decreases the computational cost of the forward simulation. This methodology of mesh generation was applied to all cases in this work.

Second-order schemes are used for spatial and temporal discretizations. The time step sizes for all cases ensure a maximum Courant-Friedrichs-Lewy (CFL) number smaller than one. For this work, the one-equation SGS TKE model as proposed by Yoshizawa and Horiuiti~\cite{yoshizawa1985} and the Wall-Adapting Local Eddy-Viscosity (WALE) model from Ducros \etal.~\cite{ducros1998wall} were used. For the forward problem, a conjugate gradient linear solver with a DIC preconditioner was used for the pressure, whereas the discretized SGS TKE was solved by a bi-conjugate gradient stabilized linear solver with a DILU preconditioner. The bi-conjugate gradient stabilized linear solver without preconditioning was applied to solve the \textit{coupled} adjoint system. Parallel computing was used for the forward simulation but not for the adjoint problem solution. The optimization itself was performed in \emph{Python} using \textit{PyFoam} to interact with the \textit{OpenFOAM} solvers.

\begin{figure}[!ht]
    \centering
    \includegraphics[width=\textwidth]{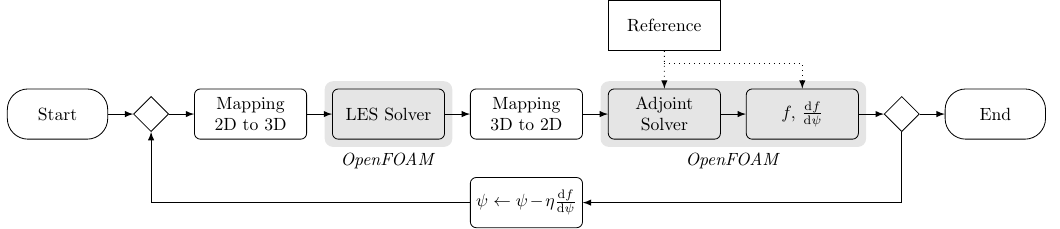}
    \caption{
        Flowchart of the optimization-based data assimilation procedure.
    }
    \label{fig:diagram_rans_da}
\end{figure}

The DA procedure is depicted in Fig.~\ref{fig:diagram_rans_da}. First, the forward problem (3D LES) is solved to obtain the forward solution $\bm U$ as a function of the parameter field $\bm \psi$. Meanwhile, time-averaging of the LES solution is performed, and the solver stops as soon as the averages converge according to Eq.~\eqref{eq:averaging convergence criterion}. Since only quasi-two-dimensional flows are considered in this work, spanwise-averaging is introduced, such that the averaged quantities are mapped on a corresponding two-dimensional mesh. Then, the time- and spanwise-averaged and converged forward system matrix $\frac{\partial \bm R}{\partial \bm U}$ is used in conjunction with the analytical evaluation of the right-hand side of Eq.~\eqref{eq:coupled_adjoint} to solve the two-dimensional adjoint system for the Lagrange multipliers $\bm \lambda$. Third, the matrix $\frac{\partial \bm R}{\partial \bm \psi}$ is constructed and, together with the regularization function and the Lagrange multipliers $\bm\lambda$, used to compute the adjoint gradient. The current parameter values and the corresponding gradient are then used in the optimization step to update the parameter field, which is mapped back onto the three-dimensional mesh.

\subsubsection{Application limits}

In general, there is no strict upper limit for the Reynolds number. As the Reynolds number increases, at least near the walls a finer computational mesh is required, which in turn raises the computational cost. This approach has so far only been applied to incompressible turbulent flows, which imposes a limitation on the Mach number, typically restricting it to Ma < 0.3.

More realistic problems will likely feature more complex geometries that incur more computational cost for the forward problem solution and they will likely lack the quasi-two-dimensional property exploited in this work. Hence, a single iteration of the parameter optimization will be much more costly.

Further, the availability of reference data that features a suitable density and number of spatially distributed locations is not guaranteed in more realistic problems and might be a limiting factor.


\subsection{Flow over periodic hills}
\label{sec:flow over periodic hills}

\begin{figure}[!ht]
    \centering
    \includegraphics[width=0.5\textwidth]{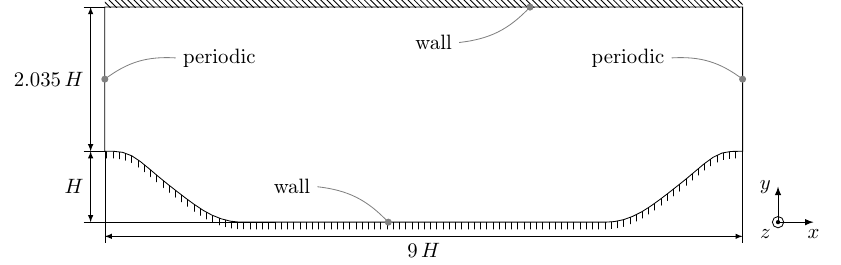}
    \caption{
            Simulation domain of the periodic hill setup. The mean flow is in $x$-direction.
            All length scales are normalized by the hill crest height $H$.
            The domain extends over a width of $4.5 \, H$ in $z$-direction, in which periodicity is assumed.
    }
    \label{fig:ph_2d}
\end{figure}

The periodic hill geometry is depicted in Fig.~\ref{fig:ph_2d}. Velocity data from highly resolved LES by Gloerfelt and Cinnella~\cite{gloerfelt19}, averaged in time and in $z$-direction, is considered as reference.
This setup features a flow at a Reynolds number of
\begin{equation}
    \mathrm{Re}
    =
    \frac{u_{b}H}{\nu}
    =
    \num{10595}
\end{equation}
based on the bulk velocity $u_{b}$ over the hill crest of height $H$ and the kinematic viscosity $\nu$.

At the upper and lower walls the no-slip boundary condition is set for the velocity and Neumann condition for the pressure and the SGS viscosity. The WALE SGS model is used to calculate the SGS viscosity.

Due to the quasi-two-dimensional flow setup (infinitely long hills in spanwise direction), a mapping onto a two-dimensional mesh with the same resolution in the $xy$-plane (but only one cell in $z$-direction) is introduced. Therefore, averaging is not only performed in time, but also in spanwise direction.

\subsubsection{Optimization}
\label{sec:optimization periodic hill fine}

The forward problem is solved on two different meshes. First, a coarse mesh with \num{74} cells in $x$-direction by \num{60} cells in $y$-direction by \num{36} cells in $z$-direction (total cell count is \num{159840}) is used. This mesh is too coarse for proper LES, particularly in near-wall regions, but the cell aspect ratio is relatively low everywhere.

\begin{figure}[!ht]
    \centering
    \begin{subfigure}[t]{0.49\textwidth}
        \centering
        \includegraphics[width=\textwidth]{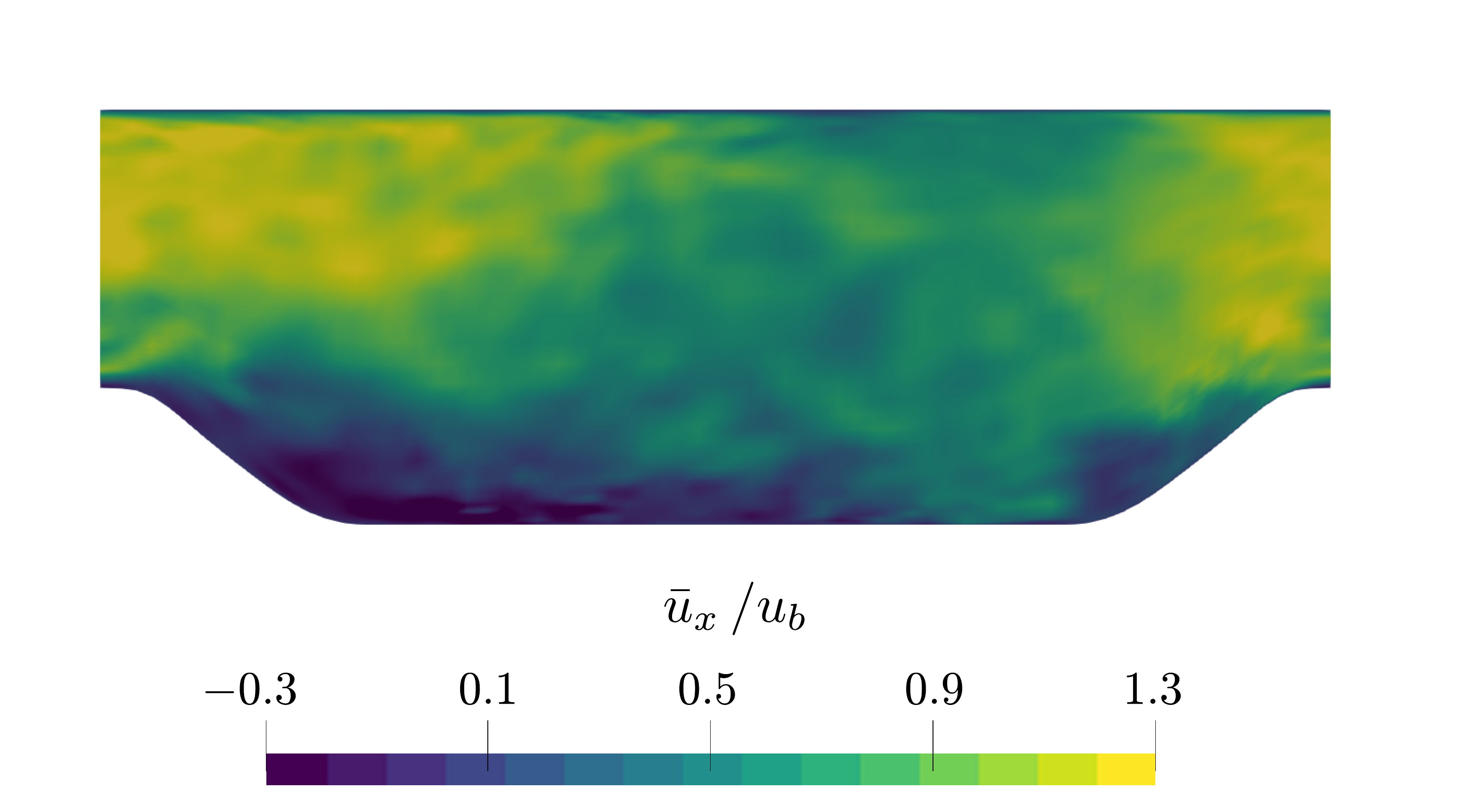}
        \caption{Instantaneous velocity snapshot ($z=0.5 \ H$).}
        \label{fig:instantaneous velocity periodic hill coarse}
    \end{subfigure}
    \hfill
    \begin{subfigure}[t]{0.49\textwidth}
        \centering
        \includegraphics[width=\textwidth]{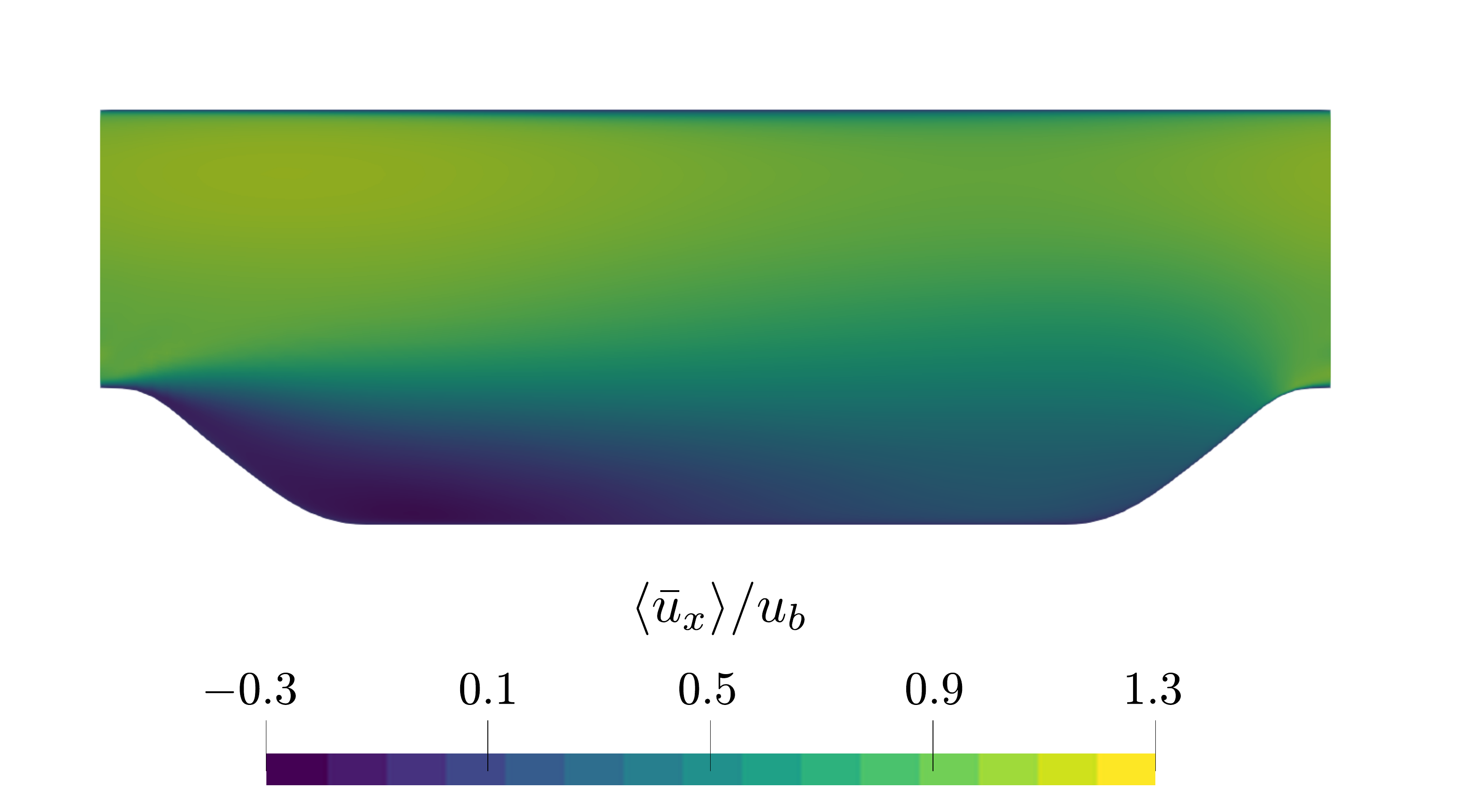}
        \caption{Time- and spanwise-averaged velocity.}
        \label{fig:averaged velocity periodic hill coarse}
    \end{subfigure}
    \caption{Streamwise velocity component of the flow over periodic hills. The LES solver ran until the convergence criterion $C_U = 0.15$ was reached.}
\end{figure}

In Fig.~\ref{fig:instantaneous velocity periodic hill coarse}, an instantaneous snapshot of the streamwise velocity component is depicted, which is taken during the run-time of the initial LES. Since the mesh is very coarse, small-scale motions are not resolved, but larger structures are clearly visible. Fig.~\ref{fig:averaged velocity periodic hill coarse} illustrates the time- and spanwise-averaged streamwise velocity component of the initial LES. As discussed in Sec.~\ref{sec:Temporal averaging of URANS equations}, the averaged quantities (here only velocity is shown) serve to construct the discrete stationary adjoint equation~\eqref{eq:urans_time_average}. Therefore, the averaging criterion (cf. Eq.~\eqref{eq:averaging convergence criterion}) needs to be set appropriately to ensure well-converged quantities such that the adjoint problem can be considered stationary.

For the optimization, reference points must be selected where averaged velocity data is assimilated. As quite extensively discussed in Brenner~\etal.~\cite{brenner_variational_2024} and Plogmann~\etal.~\cite{plogmann23}, the optimal placement of such reference data points is crucial and has a strong influence on the quality of mean flow reconstruction. Furthermore, the number of reference data points should ideally be as small as possible, as in practical applications it represents the number of (e.\,g., pointwise) measurements. The focus of this work, however, is to demonstrate that with a sufficient number of reference data points, the mean flow can be well predicted using such coarse LES.

\begin{figure}[!ht]
     \centering
     \begin{subfigure}[t]{0.49\textwidth}
         \centering
         \includegraphics[width=\textwidth]{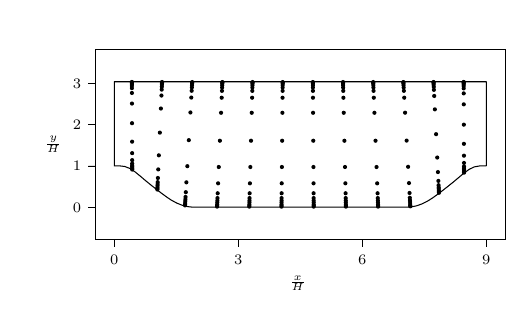}
         \caption{216 reference data points.}
         \label{fig:reference data points periodic hill fine}
     \end{subfigure}
     \hfill
     \begin{subfigure}[t]{0.49\textwidth}
         \centering
         \includegraphics[width=\textwidth]{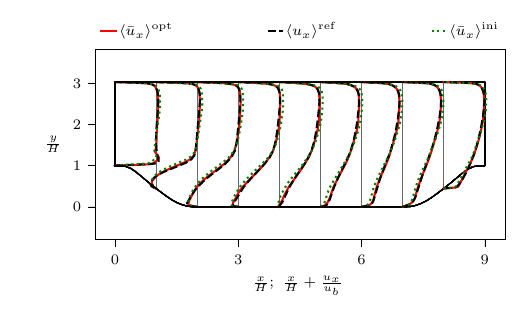}
         \caption{Profiles of the streamwise velocity component.}
         \label{fig:velocity profiles periodic hill fine}
     \end{subfigure}
     \hfill
     \begin{subfigure}[t]{0.49\textwidth}
         \centering
         \includegraphics[width=\textwidth]{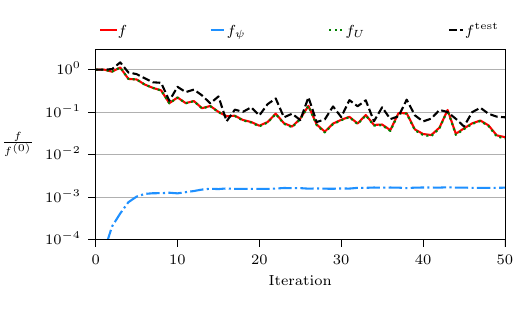}
         \caption{Test and cost function components normalized by their respective initial value $f^{(0)}$.}
         \label{fig:cost function periodic hill fine}
     \end{subfigure}
     \hfill
     \begin{subfigure}[t]{0.49\textwidth}
         \centering
         \includegraphics[width=\textwidth]{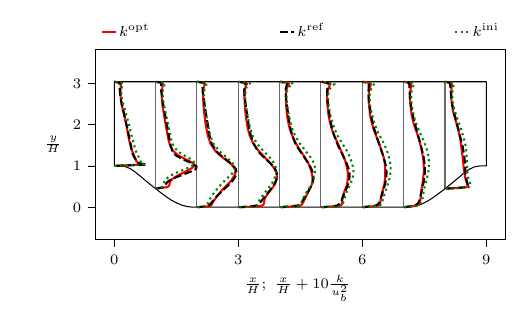}
         \caption{Profiles of the total TKE.}
         \label{fig:TKE profiles periodic hill fine}
     \end{subfigure}
    \caption{Optimization of the time- and spanwise-averaged flow over periodic hills using a coarse mesh and the WALE SGS model. Optimization step size is $\eta=\num{4e-2}$ with a maximum number of optimization steps $N_\mathrm{opt}=50$. The forward problem solver ran until the convergence criterion $C_U = 0.15$ was reached, and the regularization weight parameter was set to $C^{\mathrm{reg}}=\num{8e-7}$.}
    \label{fig:periodic hill fine results}
\end{figure}

In Fig.~\ref{fig:reference data points periodic hill fine}, the reference data distribution is shown. The density of points increases toward the walls, and an almost uniform spacing is chosen in $x$-direction. Depicted in Fig.~\ref{fig:cost function periodic hill fine}, the discrepancy part of the cost function decreases by more than one order of magnitude alongside a significant reduction of the test function due to regularization. Overall, the averaged, optimized velocity profiles shown in Fig.~\ref{fig:velocity profiles periodic hill fine} match very well with the reference. Additionally more accurate TKE predictions are obtained throughout the entire domain. Nevertheless, smaller discrepancies to the reference TKE remain in the recirculation zone or very close to the wall since the computational mesh is generally too coarse to dynamically resolve the TKE in these regions. In this work, however, the aim is to demonstrate the basic capabilities of the proposed framework rather than assessing its resolution limits.

To this end, we would like to mention that TKE data or other higher-order moments have not been assimilated. Rather, the divergence of the average residual stresses is corrected, which ultimately allows for mean flow reconstruction in conjunction with improved turbulent quantities. This topic is further elaborated on in sections \ref{sec:Anisotropic Reynolds stress reconstruction from corrective force} and \ref{sec:Replacing corrective forcing with velocity fluctuation nudging}.

\subsection{Flow around square cylinder}
\label{sec:flow around square cylinder}

\begin{figure}[!ht]
    \centering
    \includegraphics[width=0.5\textwidth]{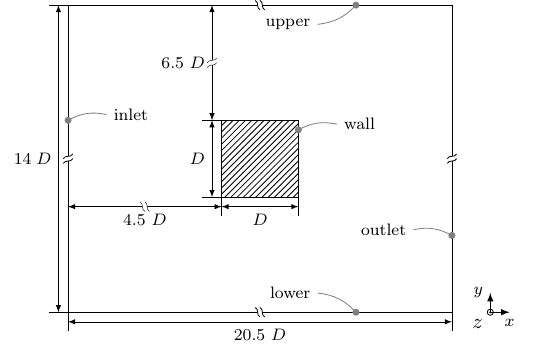}
    \caption{
            Simulation domain of the square cylinder setup with mean flow in $x$-direction.
            All length scales are normalized by cylinder width $D$, and the domain features a depth of $9.8 \, D$ in $z$-direction with periodic boundary conditions.
    }
    \label{fig:square_2d}
\end{figure}

Next, we consider the flow around a square cylinder. The forward problem is solved on a coarse (not near-wall resolving) mesh with a total of \num{139400} cells. The two-dimensional mesh (for the adjoint problem) with the same resolution in the $x$-$y$-plane therefore consists of \num{6970} cells. A sketch of the geometry and boundaries is provided in Fig.~\ref{fig:square_2d}. All length scales are expressed relative to the cylinder width $D$, which also serves as the length scale to compute the Reynolds number. The flow is analyzed for
\begin{equation*}
    \mathrm{Re}
    =
    \frac{u_{\infty}D}{\nu}
    =
    \num{22000}
    \, ,
\end{equation*}
where $u_\infty$ is the free-stream velocity. Averaged DNS data taken from \cite{trias_turbulent_2015} provides the reference.

Boundary conditions for velocity and pressure are also taken from \cite{trias_turbulent_2015}. For the SGS viscosity, Neumann conditions are applied at all boundaries. For this flow setup, the one-equation SGS TKE model is chosen. At the inlet and the cylinder wall, Dirichlet condition is applied and set to a very small value for the SGS TKE (according to \cite{trias_turbulent_2015}), and Neumann conditions are set at the outlet as well as the upper and lower boundaries.

\subsubsection{Optimization}
\label{sec:optimization cylinder}

\begin{figure}[!ht]
     \centering
     \begin{subfigure}[t]{0.49\textwidth}
         \centering
         \includegraphics[width=\textwidth]{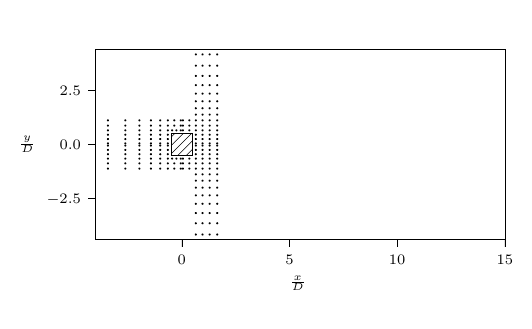}
         \caption{210 reference data points.}
         \label{fig:reference data points square cylinder}
     \end{subfigure}
     \hfill
     \begin{subfigure}[t]{0.49\textwidth}
         \centering
         \includegraphics[width=\textwidth]{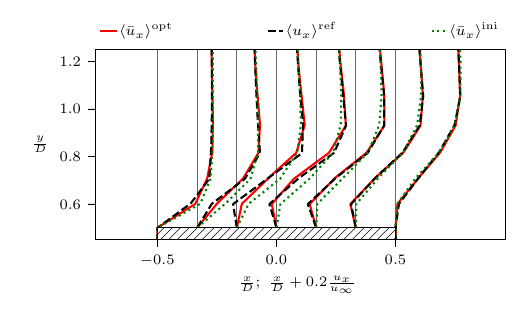}
         \caption{Profiles of the streamwise velocity component near the cylinder.}
         \label{fig:velocity profiles square cylinder near cylinder}
     \end{subfigure}
     \hfill
     \begin{subfigure}[t]{0.49\textwidth}
         \centering
         \includegraphics[width=\textwidth]{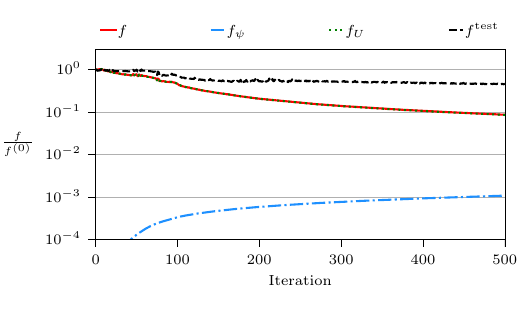}
         \caption{Test and cost function components normalized by their respective initial value $f^{(0)}$.}
         \label{fig:cost function square cylinder}
     \end{subfigure}
     \hfill
     \begin{subfigure}[t]{0.49\textwidth}
         \centering
         \includegraphics[width=\textwidth]{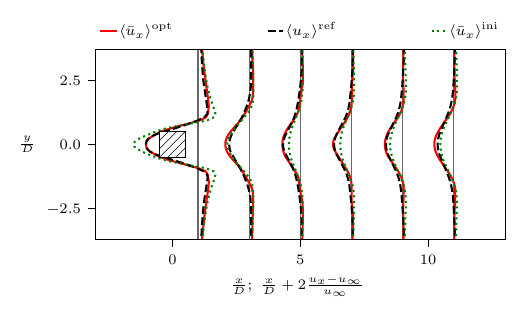}
         \caption{Profiles of the streamwise velocity component in the wake.}
         \label{fig:velocity profiles square cylinder wake}
     \end{subfigure}
    \caption{Optimization of the time- and spanwise-averaged flow around a square cylinder using a coarse mesh and the one-equation SGS TKE model. Optimization step size is $\eta=\num{1e-1}$ with a maximum number of optimization steps $N_\mathrm{opt}=500$. The forward problem solver ran until the convergence criterion $C_U = 1$ was reached, and the regularization weight parameter was set to $C^{\mathrm{reg}}=\num{1e-6}$.}
    \label{fig:square cylinder results}
\end{figure}

As visualized in Fig.~\ref{fig:reference data points square cylinder}, most of the selected reference data points are accumulated around the cylinder with additional ones along the $y$-direction in the near-wake regions. Regarding the optimization, the discrepancy part of the cost function decreases by roughly one order of magnitude, as shown in Fig.~\ref{fig:cost function square cylinder}. Moreover, the test function was only reduced by a factor of two to three. Again, we would like to mention that test data points are located in every grid cell that was not used as a source for DA.

Owing to the regularization, the averaged velocity profiles are smooth, and improvements can be observed not only around the cylinder (cf. Fig.~\ref{fig:velocity profiles square cylinder near cylinder}) but also in its wake, as illustrated in Fig.~\ref{fig:velocity profiles square cylinder wake}. Compared to DA with a URANS model, as presented in~\cite{plogmann23}, mean flow predictions also improved downstream of the cylinder, even though data was only assimilated close to the cylinder. As URANS simulations based on an eddy-viscosity model are often too dissipative, the DA using only near-cylinder reference data is not significantly improving the mean flow further downstream~\cite{plogmann23}. In contrast, LES has the advantage that large-scale motions are resolved and that only small-scale structures have to be modeled; thus, they are less dissipative. Consequently, if the objective is to improve mean flow predictions in the wake region, while observations are only available near the cylinder, it is advantageous to consider DA with LES rather than with URANS.

Since LES is unsteady, the corrective forcing term, even though it is stationary, has an impact on the flow dynamics. One essential dynamic feature of flow around a cylinder is the vortex shedding frequency. Therefore, the Strouhal number provides a good measure to assess whether the assimilation of averaged velocity data can improve the flow dynamics. We analyzed the uniformly sampled lift coefficient for roughly 65 vortex shedding cycles to obtain the Strouhal number. For the initial LES, the Strouhal number is $St^\mathrm{ini}=0.135$, while Trias~\etal.~\cite{trias_turbulent_2015} report $St^\mathrm{ref} = 0.132$ for the DNS reference simulation. After data assimilation, the optimized flow features a Strouhal number of $St^\mathrm{opt}=0.132$. Thus, while the initial LES is already close to the reference, assimilation of the averaged velocity reference data led to further improvement of the vortex shedding frequency, that is, to an almost perfect match with the reference. This behavior is also reported in~\cite{plogmann23} for URANS simulations, where after optimization, however, the match with the reference Strouhal number was not perfect.

\begin{figure}[!ht]
    \centering
    \begin{subfigure}[t]{0.49\textwidth}
        \centering
        \includegraphics[width=\textwidth]{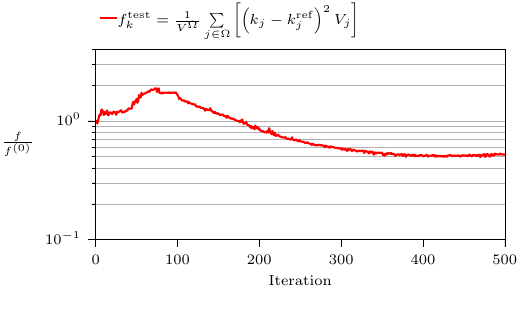}
        \caption{TKE test function normalized with its initial value $f^{(0)}$.}
        \label{fig:TKE cost function square cylinder}
    \end{subfigure}
    \hfill
    \begin{subfigure}[t]{0.49\textwidth}
        \centering
        \includegraphics[width=\textwidth]{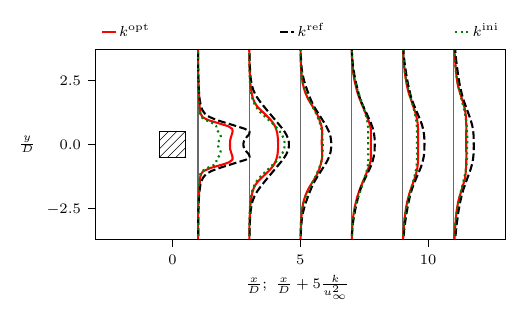}
        \caption{Profiles of the TKE in the cylinder wake.}
        \label{fig:TKE cylinder}
    \end{subfigure}
    \caption{Optimization of the total (resolved plus residual) TKE around a square cylinder.}
\end{figure}

Finally, the TKE is analyzed. For that purpose, profiles of the TKE are shown in the wake of the cylinder. A significant improvement can be observed, particularly in the near wake region where the TKE is highly underpredicted by the initial LES. For visualization purposes, a test function for the TKE, $f_k^\mathrm{test}$, is introduced, analogously to the formulation in Eq.~\eqref{eq:test function}. This function is evaluated in every grid cell of the domain. As illustrated in Fig.~\ref{fig:TKE cost function square cylinder}, the predictions of the TKE exhibit a general improvement throughout the entire computational domain as the value of $f_k^\mathrm{test}$ decreases. Again, this improvement only is a side effect since the TKE is not directly optimized, but the divergence of the average residual stresses corrected.

\section{Anisotropic Reynolds stress reconstruction from corrective force}
\label{sec:Anisotropic Reynolds stress reconstruction from corrective force}

So far, assimilation of time-averaged velocity data allowed to correct the divergence of the deviatoric part of the average residual stresses 
\begin{equation}
    \frac{\partial a_{ij}^{r}}{\partial x_{j}}
    =
    - \frac{\partial}{\partial x_j}
    \left(
    2\nu_\mathrm{sgs}\bar{S}_{ij}
    \right)
    -
    \epsilon_{ijk}\frac{\partial \psi_k}{\partial x_j}
\end{equation}
with a stationary and divergence-free force field. Hence, the divergence of the deviatoric part of a tensor can be formulated as
\begin{equation}\label{eq:divaij}
    \frac{\partial a_{ij}}{\partial x_j} = g_i 
\end{equation}
if a stationary force
\begin{equation}
    \bm{g} = \begin{bmatrix}
        g_1 \\ g_2 \\ 0
    \end{bmatrix}
\end{equation}
is uniquely determined, e.\,g. from two-dimensional data assimilation. In general, this tensor $\bm{a}$ is symmetric, i.\,e.,
\begin{align}
    a_{ij} = a_{ji} \, , ~~~ i \ne j \, ,
\end{align}
and trace-free, i.\,e.,
\begin{equation}\label{eq:tracefree}
    a_{ii} = 0 \, .
\end{equation}

Furthermore, for quasi-two-dimensional flows in the $x$-$y$ plane, $a_{13} = a_{23} = a_{31} = a_{32} = 0$ holds. Applying these properties and rewriting Eq. \eqref{eq:tracefree} as $a_{22} = - a_{11} - a_{33}$ yields the anisotropic part of the Reynolds stress tensor correction
\begin{equation}
    \bm{a} = \begin{bmatrix}
        a_{11} & a_{12} & 0 \\
        a_{12} & - a_{11} - a_{33} & 0 \\
        0 & 0 & a_{33} \\
    \end{bmatrix} \, .
\end{equation}

At this point, we would like to stress that not only for three-dimensional simulations, but also for (quasi) two-dimensional simulations, the $a_{33}$ component needs to be considered, since turbulence is inherently three-dimensional, i.\,e., the trace-free condition implies that $a_{11} + a_{22} + a_{33} = 0$. While others (e.\,g. \cite{foures14,CATO2023106054}) have attempted to compare optimized Reynolds stresses from a two-dimensional RANS simulation with Reynolds stresses obtained from three-dimensional DNS, they tied to fulfill $a_{11} + a_{22} = 0$. Since in two-dimensional RANS simulations $a_{33}$ is not available, an assumption has to be made about this tensor component (e.\,g. as done in \cite{carlucci_data-driven_2024}). In this work, however, three-dimensional LES are performed and therefore, $a_{33}$ is statistically extracted during runtime of LES. Further details are discussed in Sec.\ref{sec:Replacing corrective forcing with velocity fluctuation nudging}.

We proceed to rewrite Eq. \eqref{eq:divaij} as
\begin{align}
    \frac{\partial a_{11}}{\partial x_1} + \frac{\partial a_{12}}{\partial x_2} &= g_1 \, , \\
    \frac{\partial a_{21}}{\partial x_1} - \frac{\partial a_{11}}{\partial x_2} &= g_2 + \frac{\partial a_{33}}{\partial x_2} \, ,
\end{align}
and
\begin{equation}
    \frac{\partial a_{33}}{\partial x_3} = 0 \, ,
\end{equation}
which is always fulfilled for quasi-two-dimensional flows in the $x$-$y$ plane. Thus, we obtain two Poisson equations. For $a_{11}$ it reads
\begin{equation} \label{eq:poisson1}
    \frac{\partial^2 a_{11}}{\partial x_1 \partial x_1} + \frac{\partial^2 a_{11}}{\partial x_2\partial x_2} = -\frac{\partial g_2}{\partial x_2} + \frac{\partial g_1}{\partial x_1} - \frac{\partial^2 a_{33}}{\partial x_2\partial x_2} 
\end{equation}
and for $a_{12}$ it is
\begin{equation}\label{eq:poisson2}
    \frac{\partial^2 a_{12}}{\partial x_1 \partial x_1} + \frac{\partial^2 a_{12}}{\partial x_2\partial x_2} = \frac{\partial g_2}{\partial x_1} + \frac{\partial g_1}{\partial x_2} + \frac{\partial^2 a_{33}}{\partial x_1 \partial x_2} \, .
\end{equation}

Assuming that the force ${g}$ and the tensor component $a_{33}$ are known a priori, the two Poisson equations \eqref{eq:poisson1} and \eqref{eq:poisson2} can be solved for $a_{11}$ and $a_{12}$, respectively.

\subsection{Verification using reference data}
\label{sec:Verifciation using DNS data}

\begin{figure}[!ht]
     \centering
     \begin{subfigure}[t]{0.49\textwidth}
         \centering
         \includegraphics[width=\textwidth]{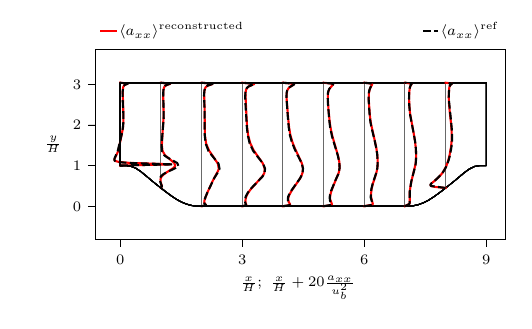}
         \caption{Profiles of $a_{xx}$.}
         \label{fig:a11}
     \end{subfigure}
     \hfill
     \begin{subfigure}[t]{0.49\textwidth}
         \centering
         \includegraphics[width=\textwidth]{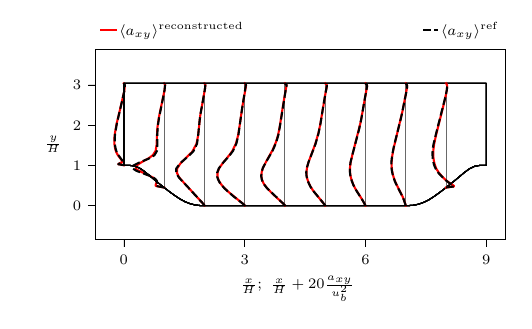}
         \caption{Profiles of $a_{xy}$.}
         \label{fig:a12}
     \end{subfigure}
     \hfill
     \begin{subfigure}[t]{0.49\textwidth}
         \centering
         \includegraphics[width=\textwidth]{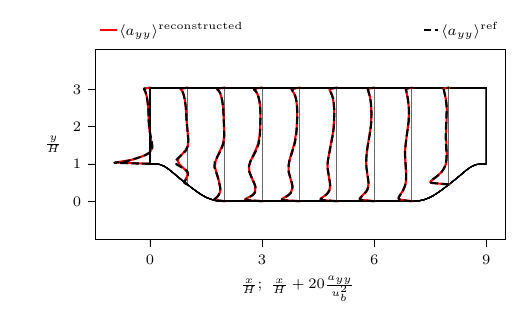}
         \caption{Profiles of $a_{yy}$.}
         \label{fig:a22}
     \end{subfigure}
     \hfill
     \begin{subfigure}[t]{0.49\textwidth}
         \centering
         \includegraphics[width=\textwidth]{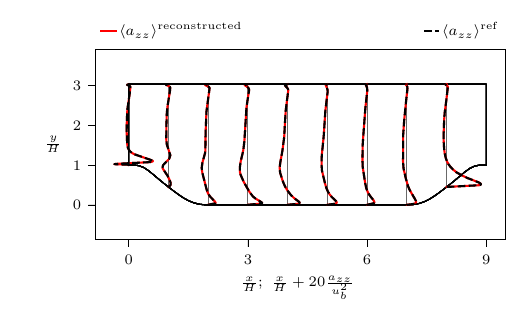}
         \caption{Profiles of $a_{zz}$.}
         \label{fig:a33}
     \end{subfigure}
    \caption{Comparison of reference and reconstructed anisotropic Reynolds stress tensor components for flow over periodic hills.}
    \label{fig:aij_comparison}
\end{figure}

To verify the proposed procedure, we assume that the anisotropic part of the Reynolds stress tensor is given (e.\,g. from highly-resolved LES reference data). By taking the divergence of this tensor, i.\,e., 
\begin{equation}
    \frac{\partial a_{ij}^\mathrm{ref}}{\partial x_j} = g_i^\mathrm{ref},
\end{equation}
we receive the reference force field. Furthermore, $a_{33}$ is known a priori and set to $a_{33} = a_{33}^\mathrm{ref}$. Solving eqs. \eqref{eq:poisson1} and \eqref{eq:poisson2} with ${g}^\mathrm{ref}$ and $a_{33}^\mathrm{ref}$ then leads to a perfect match with the anisotropic Reynolds stress components $a_{11}^\mathrm{ref}$, $a_{12}^\mathrm{ref}$ and $a_{22}^\mathrm{ref}$, as depicted in Fig. \ref{fig:aij_comparison} for the flow over periodic hills. Therefore, if the force field ${g}$ is known (e.\,g. from data assimilation) and an assumption about $a_{33}$ is made, the remaining components of the tensor can be reconstructed. 

\section{Replacing the corrective force field with forced velocity fluctuations}
\label{sec:Replacing corrective forcing with velocity fluctuation nudging}

The aim of this section is to show how a stationary force acting on the mean flow can be replaced by appropriately forcing the velocity fluctuations. First, data assimilation (see Sec. \ref{sec:methods}) must be performed to obtain a stationary correction force in the unsteady, modeled, filtered momentum Eq. \eqref{eq:rans_momentum}, which has the mean solution $\langle \bar{\bm{u}}\rangle$ obeying
\begin{equation}
\frac{\partial\langle\bar u_i\rangle}{\partial t}
+ \frac{\partial\langle\bar u_i\rangle\langle \bar u_j\rangle}{\partial x_j}
+
\frac{\partial \langle p_e\rangle}{\partial x_i}
-
\frac{\partial}{\partial x_j}\left(2\nu
\langle\bar S_{ij}\rangle
\right)
+
\underbrace{
\frac{\partial}{\partial x_j}\left(
\overbrace{-2
\left\langle\nu_\mathrm{sgs}
\bar S_{ij}
\right\rangle}^{a_{ij}^\mathrm{sgs}}
+
\overbrace{\mathrm{dev}\left(\langle\bar{u}''_{i} \bar{u}''_{j}\rangle\right)}^{a_{ij}^\mathrm{res}}\right)
-
\epsilon_{ijk}\frac{\partial \psi_k}{\partial x_j}
}_{g_i}
= 0
\end{equation}
with $\bm{a}^\mathrm{sgs}$ and $\bm{a}^\mathrm{res}$ being the sub-grid scale and resolved deviatoric (here denoted by $\mathrm{dev}(\cdot)$) parts of the Reynolds stress tensor, respectively. Analogously to Eq. \eqref{eq:rans_pressure_mod}, the isotropic contributions of the additional stress terms are also absorbed into the averaged modified pressure $\langle p_e\rangle$. After the mean flow is reconstructed through stationary forcing, the velocity fluctuations $(\bar{\bm{u}}^* - \langle\bar{\bm{u}}^*\rangle)$ are forced while solving
\begin{equation}\label{eq:LES_fluctuation_nudging}
\frac{\partial\bar  u_i^*}{\partial t}
+
\frac{\partial\bar u_i^*\bar u_j^*}{\partial x_j}
=
-
\frac{\partial p_e^*}{\partial x_i}
+
\frac{\partial}{\partial x_j}\left(2(\nu+\nu_\mathrm{sgs}^*)
\bar  S^*_{ij}
\right)
+
G_{ij}\left(
\bar {u}^*_j - \langle\bar {u}^*_j\rangle
\right)
\end{equation}
with
\begin{equation}
G_{ij}
= 
\chi
\left(a^\mathrm{res,target}_{ij} - \mathrm{dev}\left(\langle\bar{u}''^*_{i} \bar{u}''^*_{j}\rangle\right)\right) \, ,
\end{equation}
such that the same mean solution $\langle \bar{\bm{u}}^*\rangle = \langle \bar{\bm{u}}\rangle$ is obtained, where $\bm{a}^\mathrm{res,target}$ is the targeted, resolved, deviatoric part of the Reynolds stress tensor and $\chi$ is a relaxation parameter. Since the stationary force only acts on the mean flow and is subsequently replaced by fluctuation nudging, only the fluctuations are altered and the mean flow remains unchanged. Note that $\langle \bar{\bm{u}}^* \rangle$ obeys
\begin{equation}
\frac{\partial\langle\bar u_i^*\rangle}{\partial t}
+
\frac{\partial\langle\bar u_i^*\rangle\langle \bar u_j^*\rangle}{\partial x_j}
+
\frac{\partial \langle p_e^*\rangle}{\partial x_i}
-
\frac{\partial}{\partial x_j}
\left(
2\nu
\langle\bar S_{ij}^*\rangle
\right)
+
\underbrace{
\frac{\partial}{\partial x_j}
\overbrace{\left(
-2
\left\langle\nu_\mathrm{sgs}^*
\bar S^*_{ij}
\right\rangle
+
\mathrm{dev}\left(\langle\bar{u}''^*_{i} \bar{u}''^*_{j}\rangle\right)
\right)}^{a_{ij}}
}_{g_i}
= 0 \, .
\end{equation}

The necessary steps to replace the stationary force obtained from DA with a velocity fluctuation nudging term are summarized in algorithm \ref{alg:cap}.

\begin{algorithm}
\caption{Velocity fluctuation nudging procedure.}\label{alg:cap}
\begin{algorithmic}[1]
\setstretch{1.5}
\State \textbf{Require:} $\epsilon_{ijk}\frac{\partial \psi_k}{\partial x_j}$ \Comment{Perform DA to obtain stationary corrective force (see Sec. \ref{sec:methods})}

\While{$\langle \bar{\bm{u}}\rangle$ is not converged} 
    \State Solve Eq. \eqref{eq:rans_momentum}
    
    \State $a_{ij} \gets \frac{\partial}{\partial x_j}\left(-2\left\langle\nu_\mathrm{sgs}\bar S_{ij}\right\rangle+\mathrm{dev}\left(\langle\bar{u}''_{i} \bar{u}''_{j}\rangle\right)\right)$\Comment{Extract statistics}
\EndWhile

\State $a_{33}^* \gets \frac{\partial}{\partial x_3}\left(-2\left\langle\nu_\mathrm{sgs}\bar S_{33}\right\rangle+\mathrm{dev}\left(\langle\bar{u}''_{3} \bar{u}''_{3}\rangle\right)\right)$\Comment{Initialization}

\State $g_i \gets \frac{\partial a_{ij}}{\partial x_j} - \epsilon_{ijk}\frac{\partial \psi_k}{\partial x_j}$\Comment{Compute stationary force for anisotropic Reynolds stress reconstruction}
    
\While{$\frac{\partial}{\partial x_j}\left(
-2
\left\langle\nu_\mathrm{sgs}^*
\bar S^*_{ij}
\right\rangle
+
\mathrm{dev}\left(\langle\bar{u}''^*_{i} \bar{u}''^*_{j}\rangle\right)
\right)
\neq g_i
$}\Comment{Check convergence according to Eq. \eqref{eq:g_test}}
    
    \State $\bm{a} \gets \mathrm{Poisson}\left(\bm{g}, a_{33}^*\right)$ \Comment{Solve eqs. \eqref{eq:poisson1} and \eqref{eq:poisson2} to get corrected anisotropic Reynolds stresses}
    
    \State $a_{ij}^\mathrm{res,target} \gets a_{ij} + 2\left\langle\nu_\mathrm{sgs}^* \bar{S}^*_{ij}\right\rangle$ \Comment{Update resolved, targeted, deviatoric part of Reynolds stress tensor}

    \State $G_{ij} \gets \chi
    \left(a^\mathrm{res,target}_{ij} - \mathrm{dev}\left(\langle\bar{u}''^*_{i} \bar{u}''^*_{j}\rangle\right)\right)$ \Comment{If 2D, only force $G_{11}$, $G_{12}$, $G_{21}$ and $G_{22}$}

    \While{$\mathrm{dev}\left(\langle\bar{u}''^*_{i} \bar{u}''^*_{j}\rangle\right)
    \neq a^\mathrm{res,target}_{ij}
    $}
        \State Solve Eq. \eqref{eq:LES_fluctuation_nudging}
        \State $a_{33}^* \gets \frac{\partial}{\partial x_3}\left(-2\left\langle\nu_\mathrm{sgs}^*\bar S^*_{33}\right\rangle+\mathrm{dev}\left(\langle\bar{u}''^*_{3} \bar{u}''^*_{3}\rangle\right)\right)$\Comment{Extract statistics}
        
    \EndWhile
\EndWhile
\end{algorithmic}
\end{algorithm}

Notably, the corrective force only acts on the two-dimensional mean flow and hence, the tensor component $a_{33}$ is not subject to fluctuation nudging (see line 11 in algorithm \ref{alg:cap}). Therefore, $a_{33}^*$ is statistically extracted during the solution of the forward problem, i.\,e., during the solution of Eq. \eqref{eq:LES_fluctuation_nudging}, and subsequently used to calculate the corrected anisotropic part of the Reynolds stress tensor.

\subsection{Flow over periodic hills}   
\label{sec:flow over periodic hills_2}

\begin{figure}[!ht]
     \centering
     \begin{subfigure}[t]{0.49\textwidth}
         \centering
         \includegraphics[width=\textwidth]{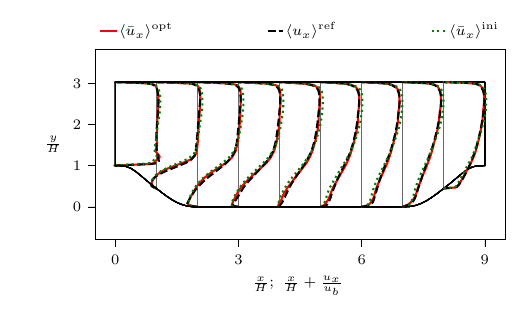}
         \caption{Profiles of the streamwise velocity component.}
         \label{fig:velocity profiles periodic hill anisotropic}
     \end{subfigure}
     \hfill
     \begin{subfigure}[t]{0.49\textwidth}
         \centering
         \includegraphics[width=\textwidth]{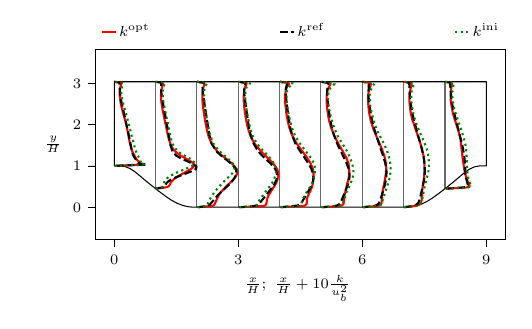}
         \caption{Profiles of the total TKE.}
         \label{fig:TKE profiles periodic hill anisotropic}
     \end{subfigure}
     \hfill
     \begin{subfigure}[t]{0.49\textwidth}
         \centering
         \includegraphics[width=\textwidth]{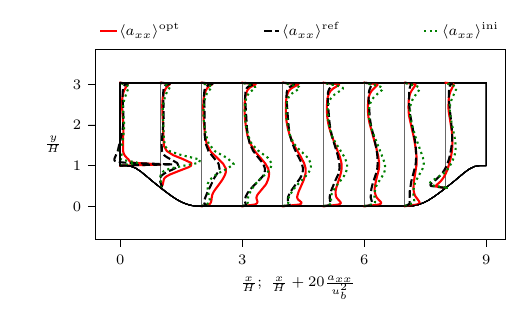}
         \caption{Profiles of $a_{xx}$.}
         \label{fig:a11 periodic hill anisotropic}
     \end{subfigure}
     \hfill
     \begin{subfigure}[t]{0.49\textwidth}
         \centering
         \includegraphics[width=\textwidth]{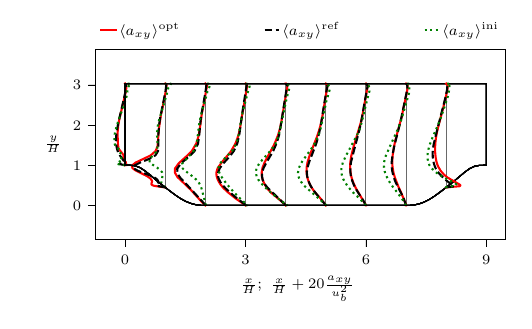}
         \caption{Profiles of $a_{xy}$.}
         \label{fig:a12 periodic hill anisotropic}
     \end{subfigure}
     \hfill
     \begin{subfigure}[t]{0.49\textwidth}
         \centering
         \includegraphics[width=\textwidth]{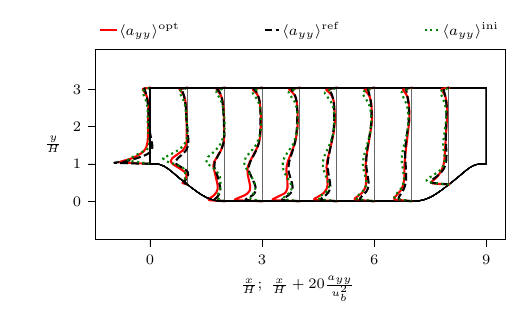}
         \caption{Profiles of $a_{yy}$.}
         \label{fig:a22 periodic hill anisotropic}
     \end{subfigure}
     \hfill
     \begin{subfigure}[t]{0.49\textwidth}
         \centering
         \includegraphics[width=\textwidth]{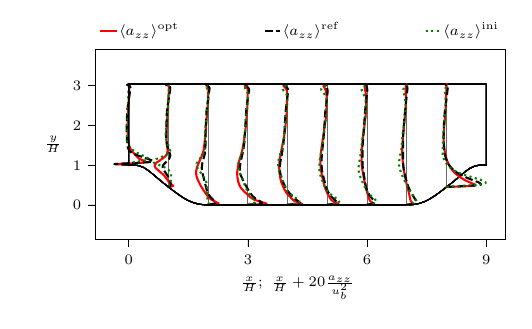}
         \caption{Profiles of $a_{zz}$.}
         \label{fig:a33 periodic hill anisotropic}
     \end{subfigure}
    \caption{Mean flow, TKE and anisotropic Reynolds stress components after velocity fluctuation nudging for flow over periodic hills using a coarse mesh. Stationary corrective forcing was switched off and fluctuation nudging turned on after $500 H/u_b$ until $1000 H/u_b$. The relaxation parameter was set to $\chi=400$. Convergence of the velocity fluctuation nudging algorithm was achieved after five iterations, as shown in Fig. \ref{fig:cost function g periodic hills}.}
    \label{fig:periodic hill anisotropic reynolds stress}
\end{figure}

Figure~\ref{fig:velocity profiles periodic hill anisotropic} shows that the mean flow remains in the optimized state after fluctuation nudging. The situation is similar for the TKE, which continues to show an improvement throughout the entire domain, but remains inaccurate close to the walls (see Fig. \ref{fig:TKE profiles periodic hill anisotropic}). As already mentioned several times, the TKE is not part of the optimization, but is only shown here for completeness. Figures \ref{fig:a11 periodic hill anisotropic} to \ref{fig:a33 periodic hill anisotropic} depict the anisotropic parts of the Reynolds stress components. In principle, a clear improvement can be seen for all tensor components. However, for the components $a_{xx}$ and $a_{yy}$, the improvement near the walls only is moderate. Again, it is conceivable that the insufficient mesh resolution hinders a higher accuracy. The effect of the computational mesh on the velocity fluctuation nudging term is further investigated in Sec.\ref{sec:flow around square cylinder_2}. At this point it is important to note once again that the component $a_{zz}$ was not directly corrected by the force determined from data assimilation. Rather, the optimization of $a_{zz}$ is a positive side effect of the velocity fluctuation scaling through nudging of the other stress tensor components.

The discrepancy between the simulation results and the reference data in the anisotropic Reynolds stresses stems from the underdetermined nature of the inverse problem, which results in a non-unique correcting force. This is due to the fact that the number of reference data points (measurements) does not match the number of parameters (number of grid cells). Consequently, it is likely that the corrective force will be slightly different if a different distribution of reference data points is chosen. Unfortunately, this phenomenon cannot be completely avoided for sparse data assimilation. However, for other applications (e.\,g. \cite{luther2025nonlinearsuperstencilsturbulencemodel}) where a force term is uniquely determined, it is conceivable that this problem will not occur.

\subsection{Flow around square cylinder}
\label{sec:flow around square cylinder_2}

\begin{figure}[!ht]
     \centering
     \begin{subfigure}[t]{0.49\textwidth}
         \centering
         \includegraphics[width=\textwidth]{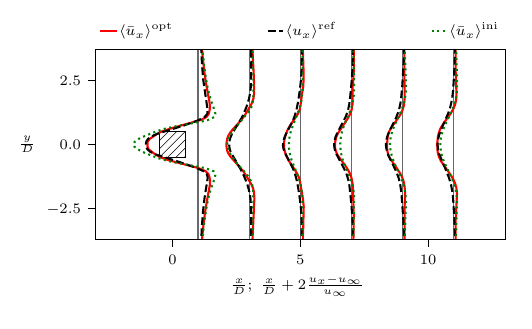}
         \caption{Profiles of the streamwise velocity component.}
         \label{fig:velocity profiles SC anisotropic}
     \end{subfigure}
     \hfill
     \begin{subfigure}[t]{0.49\textwidth}
         \centering
         \includegraphics[width=\textwidth]{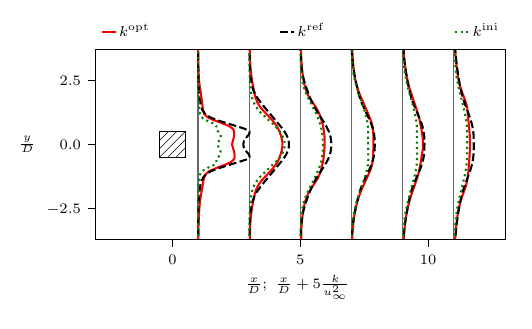}
         \caption{Profiles of the total TKE.}
         \label{fig:TKE profiles SC anisotropic}
     \end{subfigure}
     \hfill
     \begin{subfigure}[t]{0.49\textwidth}
         \centering
         \includegraphics[width=\textwidth]{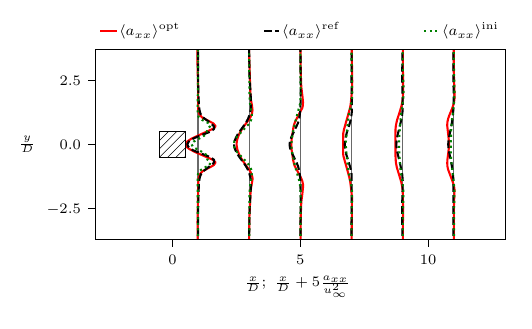}
         \caption{Profiles of $a_{xx}$.}
         \label{fig:a11 SC anisotropic}
     \end{subfigure}
     \hfill
     \begin{subfigure}[t]{0.49\textwidth}
         \centering
         \includegraphics[width=\textwidth]{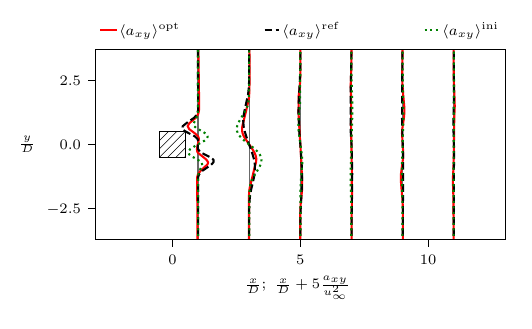}
         \caption{Profiles of $a_{xy}$.}
         \label{fig:a12 SC anisotropic}
     \end{subfigure}
     \hfill
     \begin{subfigure}[t]{0.49\textwidth}
         \centering
         \includegraphics[width=\textwidth]{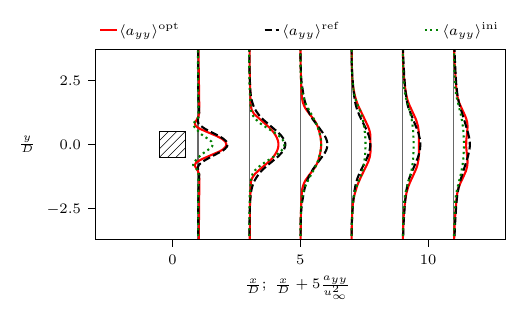}
         \caption{Profiles of $a_{yy}$.}
         \label{fig:a22 SC anisotropic}
     \end{subfigure}
     \hfill
     \begin{subfigure}[t]{0.49\textwidth}
         \centering
         \includegraphics[width=\textwidth]{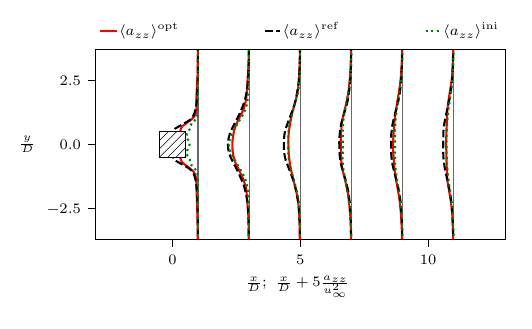}
         \caption{Profiles of $a_{zz}$.}
         \label{fig:a33 SC anisotropic}
     \end{subfigure}
    \caption{Mean flow, TKE and anisotropic Reynolds stress components after velocity fluctuation nudging for flow around a square cylinder. Stationary corrective forcing was switched off and fluctuation nudging turned on after $500 D/u_\infty$ until $1500 D/u_\infty$.The relaxation parameter was set to $\chi=150$. Convergence of the velocity fluctuation nudging algorithm was achieved after five iterations, as shown in Fig. \ref{fig:cost function g SC}.}
    \label{fig:SC anisotropic reynolds stress}
\end{figure}

Analogous to the flow over periodic hills, the mean flow and TKE are also optimized for the flow around a square cylinder after fluctuation nudging (see Figs. \ref{fig:velocity profiles SC anisotropic} and \ref{fig:TKE profiles SC anisotropic}). For the anisotropic parts of the Reynolds stresses, there is a significant improvement compared to the initial uncorrected LES simulation, especially in the near wake of the cylinder (see Figs. \ref{fig:a11 SC anisotropic} to \ref{fig:a33 SC anisotropic}).

\begin{figure}[!ht]
     \centering
         \includegraphics[width=0.49\textwidth]{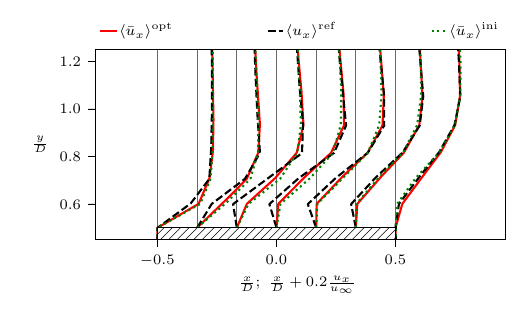}
     \caption{Profiles of the streamwise velocity component near the cylinder.}
     \label{fig:velocity profiles SC anisotropic near wall}
\end{figure}

Note that the flow, especially the wall-shear stresses, is heavily under-resolved by the computational mesh. Therefore, the divergence-free force field does not only correct the physical discrepancies introduced by the sub-grid scale model, but also accounts for the numerical errors introduced by the insufficient resolution of the computational mesh. As a result, velocity fluctuations are not resolved near the wall and hence, the velocity fluctuation nudging has no effect in this region, as can be seen in Fig. \ref{fig:velocity profiles SC anisotropic near wall}. 

One way to solve this problem is to keep the corrective force term near the wall and only rely on fluctuation nudging away from the wall. A near-wall damping through an elliptic relaxation term (e.\,g. as described in \cite{manceau}) that is multiplied with the corrective force would be conceivable.


\section{Conclusions and outlook}
\label{sec:conclusion}

Turbulent dispersion is a crucial phenomenon in fluid dynamics where the chaotic motion leads to enhanced mixing and transport of substances, such as pollutants, or heat. Since most URANS simulations rely on eddy-viscosity models, which are often too dissipative, they lack the ability to accurately predict turbulent dispersion. LES, on the other hand, resolves some of the turbulent scales and is thus more predictive.

In this work we proposed a three-dimensional variational assimilation of sparse time-averaged velocity reference data into LES by means of a stationary divergence-free forcing term in the respective momentum equation. The stationary discrete adjoint method was leveraged to compute the cost function gradient at a low computational cost. To do so, the filtered Navier--Stokes equations were time-averaged and subsequently used to construct a stationary adjoint equation. Making use of our efficient semi-analytical approach to compute the cost function gradient and the gradient-based optimizer for the parameter update, we demonstrated that for highly under-resolved LES, mean velocity predictions can significantly be improved. In particular, we investigated two types of flows, namely, the flow over periodic hills with a broadband frequency spectrum and the flow around a square cylinder with periodic vortex shedding. We also demonstrated that the stationary corrective forcing in the instantaneous LES equations improved the simulated flow dynamics. After optimization, the Strouhal number, which characterizes the periodic vortex shedding frequency, is in perfect agreement with the Strouhal number from the high-fidelity DNS.

Furthermore, the anisotropic Reynolds stress predictions improved after the assimilation of sparse time-averaged velocity reference data for both flow configurations. This was achieved by calculating the corrected anisotropic Reynolds stresses from the corrective stationary force and subsequently replacing it with a nudging term that rescales the velocity fluctuations.

Future work should investigate the optimal placement of reference data points, aiming at minimizing the number of required observations while ensuring accurate mean flow reconstruction and effective dynamic flow control. So far, only time-averaged velocity reference data has been assimilated and TKE prediction enhancement was limited, since only the deviatoric part of the Reynolds stress tensor was optimized through a divergence-free force field. Similarly, TKE reference data can be assimilated by deriving a TKE transport equation and optimizing a force term within this equation. Additionally, the resolution of the computational mesh in combination with other sub-grid scale models should be further studied to assess its influence on the DA procedure.


\section*{Data availability}

The computational codes employed in this study are publicly accessible in \cite{github}.

\section*{Acknowledgments}
\label{sec:acknowledgments}

The authors would like to thank Jonas Luther, Pasha Piroozmand, and Arthur Couteau for many helpful discussions.


\appendix
\setcounter{figure}{0}

\section{Computational cost of discrete adjoint method}
\label{app:Computational cost discrete adjoint method}

The use of a scale-resolving, three-dimensional forward solver makes an efficient adjoint solution crucial to keep the total computational cost relatively low, since several optimization steps are needed. Hence, the computational cost of the adjoint solver should be a fraction of the one from the forward solver.

Brenner~\cite{brennerthesis24} performed a preliminary investigation into the computational cost associated with the adjoint optimization process. The majority of the data assimilation cost is related to the forward problem solution and adjoint gradient computation performed in every optimization step. 

To evaluate the computational cost of the discrete adjoint approach~\cite{brenner22}, employed in this work, we compare the time of solving the forward problem (LES) until convergence of the time-averaged velocity is reached with the time of solving the adjoint problem. The analysis is performed for the periodic hill setup presented in Sec.~\ref{sec:flow over periodic hills}. The Linux \texttt{time} command was used to measure the CPU times of the forward (LES) and adjoint solvers.

The analyses presented in this appendix are run on an 9th generation node of the ETH Zurich Euler cluster (Euler IX), in particular, on an AMD EPYC 9654 CPU node. The forward solver runs in parallel mode on 128 CPU cores with 131072 MB of memory allocated, which is considerably more than necessary. The adjoint solver runs in serial mode. 

In total, 50 optimization steps are performed (cf. Fig.~\ref{fig:cost function periodic hill fine}), where the average forward solver run lasts 344.78 s and the average adjoint solver runs for 0.191 s. The adjoint problem solution thus accounts for 0.055\,\% of the total (forward problem and adjoint problem) computational cost. We would like to highlight that the computational cost is so low because the method is approximate and semi-analytical, which does not require computationally expensive approaches for the computation of the partial derivatives, such as automatic differentiation (AD) or finite differences (FD). Kenway \etal.~\cite{KENWAY2019100542} formulated a discrete adjoint approach in \textit{OpenFOAM} to perform aerodynamic shape optimization using a gradient-descent optimization. They offer two different implementations in their code. The first one is based on FD, where the computation of the partial derivatives is accelerated by a graph-coloring solver. The second implementation relies on overloading AD, which is Jacobian-free and is faster but more memory consuming than the FD implementation. However, in serial mode, the computational cost of the adjoint solution is 4.7 times higher than the forward problem solution. In other words, the adjoint solver accounts for 82.52\,\% of the total time spent between forward and adjoint solutions. He \etal.~\cite{HE2018285} shows that in parallel mode, the computation of the adjoint solution is accelerated. Nevertheless, the computational cost still is in the same order of magnitude as the one of the forward solver.

In conclusion, the discrete adjoint approach employed in this work is of approximate nature but its efficiency is unparalleled.

\section{Averaging convergence criterion}

We introduce a global measure for the change of an averaged field, e.\,g., of the velocity between time steps $(i-1)$ and $(i)$ as
\begin{equation}
\label{eq:averaging convergence criterion}
    C_U = \int_\Omega \frac{\delta \lVert \langle \bar{u} \rangle \rVert}{\delta t} \mathrm{d}V = \int_\Omega \frac{\lVert \langle \bar{u} \rangle ^{(i)} - \langle \bar{u} \rangle ^{(i-1)} \rVert}{\Delta t ^{(i)}} \mathrm{d}V \approx \sum_k \frac{\lVert \langle \bar{u} \rangle _k ^{(i)} - \langle \bar{u} \rangle _k ^{(i-1)} \rVert}{\Delta t ^{(i)}} V_k
\end{equation}
with the  average velocity vector $\langle \bar{u}\rangle_k$ in cell $k$, the vector norm $\lVert \cdot \rVert$, the time step size $\Delta t$, and the cell volume $V_k$.

\section{Velocity fluctuation nudging convergence criterion}
\label{app:Additional results for flow over periodic hills}

To evaluate the termination criterion for the fluctuation nudging algorithm \ref{alg:cap}, a test function for $\bm{g}$ is formulated as
\begin{equation}
\label{eq:g_test}
    f_g^\mathrm{test} = \frac{1}{V^{\Omega}}\sum\limits_{j\in\Omega}\left[\sum\limits_{k\in\left\{x,y,z\right\}}\left(\frac{\partial}{\partial x_l}\left(-2\left\langle\nu_{\mathrm{sgs},j}^*\bar S^{*}_{kl,j}\right\rangle+{\mathrm{dev}}\left(\langle\bar{u}^{\prime\prime *}_{k,j} \bar{u}^{\prime\prime *}_{l,j}\rangle\right)\right)-g_{k,j}\right)^{2}V_j\right] \, .
\end{equation}

\begin{figure}[!ht]
     \centering
         \includegraphics[width=0.49\textwidth]{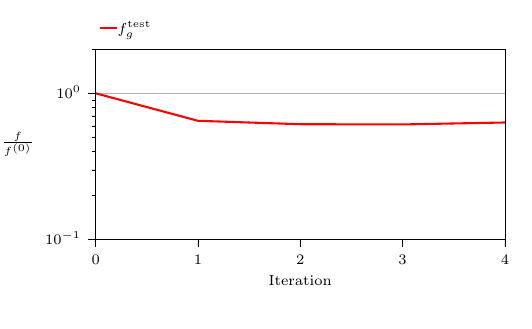}
     \caption{Test function $f_g^\mathrm{test}$ for flow over periodic hills normalized with its initial value $f^{(0)}$.}
     \label{fig:cost function g periodic hills}
\end{figure}

As can be seen in Figs. \ref{fig:cost function g periodic hills}, $f_g^\mathrm{test}$ converges after very few iterations.

\begin{figure}[!ht]
     \centering
         \includegraphics[width=0.49\textwidth]{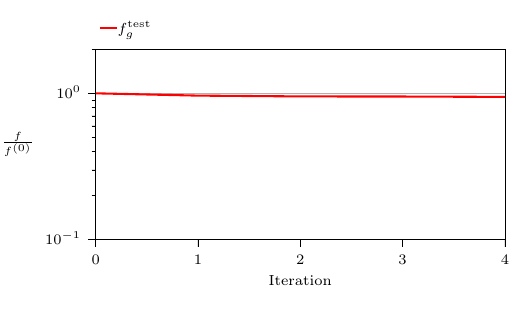}
     \caption{Test function $f_g^\mathrm{test}$ for flow around a square cylinder normalized with its initial value $f^{(0)}$.}
     \label{fig:cost function g SC}
\end{figure}


\end{document}